\documentclass[aps,pre,twocolumn]{revtex4}
\usepackage{amsmath,bm,epsfig,,subfigure}
\usepackage{color}
\usepackage[normalem]{ulem}
\def\Fbox#1{\vskip1ex\hbox to 8.5cm{\hfil\fboxsep0.3cm\fbox{%
  \parbox{8.0cm}{#1}}\hfil}\vskip1ex\noindent}  

\newcommand{\B}[1]{{\bm{#1}}}
\newcommand{\C}[1]{{\mathcal{#1}}}    
\let \= \equiv \let\*\cdot \let\~\widetilde \let\-\overline

\let \= \equiv \let\*\cdot \let\~\widetilde \let\^\widehat \let\-\overline

\begin{document}
\title{Shear localization in 3-Dimensional Amorphous Solids}
\author{Ratul Dasgupta$^1$, Oleg Gendelman$^2$, Pankaj Mishra$^1$, Itamar Procaccia$^1$ and Carmel A.B.Z. Shor$^1$}
\affiliation{$^1$ Dept. of Chemical Physics, The Weizmann Institute of Science, Rehovot 76100, Israel\\
$^2$ Faculty of Mechanical Engineering, Technion, Haifa 32000, Israel.}
\begin{abstract}
In this paper we extend the recent theory of shear-localization in 2-dimensional amorphous solids
to 3-D. In 2-D the fundamental instability of shear-localization is related
to the appearance of a line of displacement quadrupoles, that makes an angle of 45$^\circ$ with the
principal stress axis. In 3-D the fundamental plastic instability is also explained by the formation of a lattice of anisotropic elastic inclusions. In the case of pure external shear stress, we demonstrate that this is a 2-dimensional triangular lattice of similar elementary events.  It is shown that this lattice is arranged on a plane, that,  similarly to the 2-D case, makes an angle of  45$^\circ$ with respect to the principal stress axis. This solution is energetically favorable only if the external strain exceeds a yield-strain value, which is determined by the strain parameters of the elementary events and the Poisson ratio. The predictions of the theory are compared to numerical simulations and very good agreement is observed.
\end{abstract}
\maketitle

\section{introduction}

Shear banding occurs in a solid material subject to external homogeneous strain (be it simple shear, compression or extension). In some range of the external strain, the material undergoes an instability and the strain becomes localized in narrow bands. This stress localization is usually followed by the material breaking. The lowest value of the external strain $\gamma$ for which the phenomenon may appear is referred to as the yield strain $\gamma_{_Y}$. It is quite desirable  to provide a comprehensive theory for the computation of $\gamma_{_Y}$ from the first principles, especially for the quite challenging case of amorphous solids. Such a theory was obtained recently for the case of 2-dimensional amorphous solids \cite{12DHP,13DHP}. The goal of this paper is to extend this theory to 3-dimensions.

Below we consider the model of amorphous solids obtained by quenching a supercooled liquid, which consists of $N$ particles in
a volume $V$. The total potential energy of the system is denoted as $U(\B r_1,\B r_2, \cdots \B r_N)$ where
$\{ \B r_i\}_{i=1}^N$ is the set of positions of the particles.
The shear banding is easier to understand at $T=0$ and for quasi-static strain. If one develops the theory at these conditions, it is possible to go on and to determine the yield strain at any temperature and any shear rate \cite{13DJHP}.
In 2-dimensional amorphous solids it is now well known that the elementary plastic instabilities are seen when an
eigenvalue $\lambda_P$ of the Hessian matrix
\begin{equation}
H_{ij} \equiv \frac{\partial^2 U(\B r_1,\B r_2, \cdots \B r_N)}{\partial \B r_i \partial \B r_j} \ .
\end{equation}
hits zero through a saddle-node bifurcation. When this happens, the associated eigenfunction is localized on a
two-dimensional quadrupolar structure that describes the displacement field associated with the plastic
event: some particles displace into the core of the quadrupole and some displace outward from the core,
releasing locally some stress and some energy. This quadrupole-like structure can be modeled by the displacement field of a circular Eshelby inclusion that had been strained to an ellipse. Below we demonstrate that this way of modeling can be extended for the case of  3-dimensions.
Due to large structural fluctuations typical for the amorphous solid, such elementary plastic event can occur at any
value of the external strain $\gamma \ne 0$. However, the single event is not enough to form the shear band.

When the external strain increases, and particularly when $\gamma$ reaches the yield strain $\gamma_{_Y}$, a completely new scenario comes up. A highly correlated line of such interacting quadrupoles appears simultaneously and causes a drop in energy that depends on the system size. The
displacement field lines associated with this set of quadrupoles, are now
connected globally, and the strain is localized in a narrow band. It had been shown in \cite{12DHP,13DHP} that this particular arrangement of quadrupoles is a minimal energy solution in which all the quadrupoles
are ``in phase" and are aligned at 45$^\circ$ to the principal stress axis.

Even in view of the above results, it is still unclear what is the fundamental plastic instability in 3-dimensions, let alone what is the nature
of the shear-banding instability in 3-D. In this paper we will provide a theory that will clarify these
issues when the system undergoes the simple  shear strain in a plane ($\hat {\B x}-\hat {\B y}$).

The structure of the paper is as follows. In Sect.~\ref{simulations} we present numerical simulations that indicate very strongly that
also the 3-dimensional systems under the shear-strain exhibit elementary plastic events that can be faithfully modeled
by Eshelby inclusions. These inclusions occur to be spherical inclusions that have been strained in only two dimensions, with one outgoing and one incoming directions. The third direction (the $\hat {\B z}$ direction) remains neutral. It is physically reasonable, if one recalls  the geometry of an $\hat {\B x}-\hat {\B y}$ external strain, which creates no net forces in  $\hat {\B z}$-direction.
 In Sect.~\ref{theory} we present an explicit solution of Navier-Lam\`e equations for such inclusions, to provide the analytic form of the displacement field associated with elementary plastic events
in 3-D. In Sect.~\ref{numexloc} we return to the numerical simulations to examine carefully the fist shear localization event that occurs when $\gamma=\gamma_{_Y}$. This event is associated with the sub-extensive energy drop. It means that such an event must contain a {\em density} of the elementary events. Then, it is important to understand how this density is organized in space. The numerics indicates that the instability at $\gamma=\gamma_{_Y}$ is associated with the tight structuring of $\C N$ elementary plastic events on the triangular lattice in a single plane at 45$^\circ$ with the principal stress axis. To understand this fascinating phenomenon, in Sect.~\ref{energysect} we compute the energy necessary to create $\C N$ such inclusions in a given order. We show that the minimal energy for a given density is indeed obtained when these inclusions are arranged
on a plane that is in 45$^\circ$ with the principal stress and form the triangular lattice. In Sect.~\ref{compare} we compare the predictions of the analytic theory to
numerical simulations. Finally in Sect.~\ref{gamy} we compute analytically
the yield strain where the first shear localization event may take place. Sect. \ref{conclusions} offers a summary of the paper and some concluding remarks.

\section{Numerical Experiments: the Elementary Plastic Event}
\label{simulations}

For the purposes of our numerical experiments we chose the extensively
studied Kob-Andersen model \cite{95KA} as a glass former. This model employs a 80:20 binary mixture of Lennard-Jones particles
referred to as A and B at a total density $\rho=\rho_A+\rho_B=1.22$. The A and B particles interact via the potential
\begin{eqnarray}
U_{ij}(r_{ij}) &=& 4\epsilon_{ij}\Big[\Big(\frac{\sigma_{ij}}{r_{ij}}\Big)^{12} - \Big(\frac{\sigma_{ij}}{r_{ij}}\Big)^{6} + A_0 \nonumber\\ &+& A_2\Big(\frac{r_{ij}}{\sigma_{ij}}\Big)^2 + A_4\Big(\frac{r_{ij}}{\sigma_{ij}}\Big)^4 + A_6\Big(\frac{r_{ij}}{\sigma_{ij}}\Big)^6\Big] \ , \label{Uij}
\end{eqnarray}
The parameters $\epsilon_{ij}, \sigma_{ij}$ were chosen to agree with those in Ref. \cite{04VBB}. The parameters
$A_0, A_2, A_4,A_6$ are chosen to cut-off the potential at distance 2.5$\sigma_{AA}$ with two smooth derivatives.

The solid samples for our simulations were obtained by cooling a liquid equilibrated at $T=0.5$ using standard
molecular dynamics and then quenching the resulting fluid to $T=0.001$ at a rate of 6.4$\times 10^{-6}$ in LJ units.
The results shown in this article were taken from $N=64000$ and $N=85184$, although similar results were obtained for other system sizes ranging from 2000  to 144000 particles.

As discussed  above, we considered the theory for the case of external simple shear because for this case the strain tensor is traceless and thus simplifying some of the theoretical expressions. Applying an external shear in atherrmal quasi-static conditions (AQS), one discovers that the response of an amorphous solids to a small increase in the external shear strain $\delta\gamma$ (we drop tensorial indices for simplicity) is composed of two contributions. The first is the affine response which simply follows the imposed shear, such that the particles positions $\B r_i=(x_i,y_i,z_i)$ change via
\begin{eqnarray}
x_i &\to& x_i+\delta\gamma \,y_i \equiv x'_i ,\nonumber\\
y_i &\to& y_i \equiv y'_i , \\
z_i&\to& z_i \equiv z'_i.
\end{eqnarray}
 This affine response results in non-zero forces between the particles in an amorphous solid. These non-affine forces  are relaxed by the non-affine response $\B u_i$ which returns the system to mechanical equilibrium. Thus in total $\B r_i\to \B r'_i+\B u_i$. The non-affine response $\B u_i$ satisfies the exact (and model independent) differential equation of the form \cite{ML,11HKLP}
\begin{equation}
\frac{d\B u_i}{d\gamma} = -H^{-1}_{ij} \Xi_j
\end{equation}
where $\Xi_i\equiv
 \frac{\partial^2 U(\B r_1, \B r_2, \cdots \B r_n;\gamma)}{\partial \gamma \partial \B r_i}$ is known as the non-affine force.
 The inverse of the Hessian matrix is evaluated after the removal of all Goldstone modes (if they exist).
 \subsection{Stress-strain and energy-strain curves}
\begin{figure}
\includegraphics[scale = 0.4]{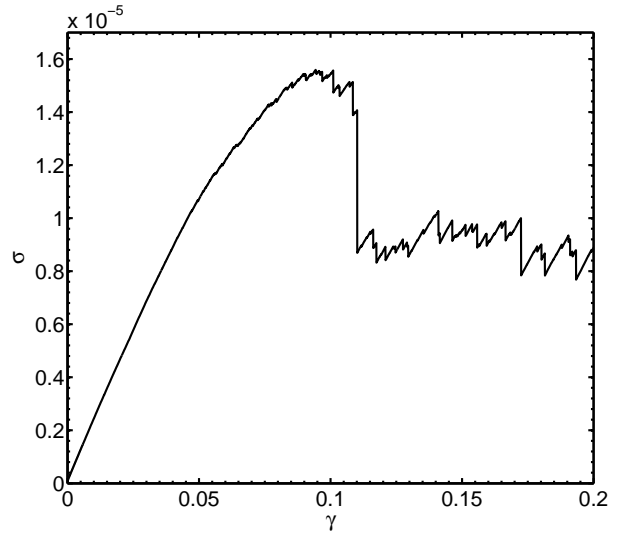}
\caption{Typical stress $\sigma$ vs strain $\gamma$ in an AQS protocol for our 3-D amorphous solid (N=85184).  }
\label{sigvsgam}
\end{figure}

\begin{figure}
\vskip 0.5 cm
\includegraphics[scale = 0.40]{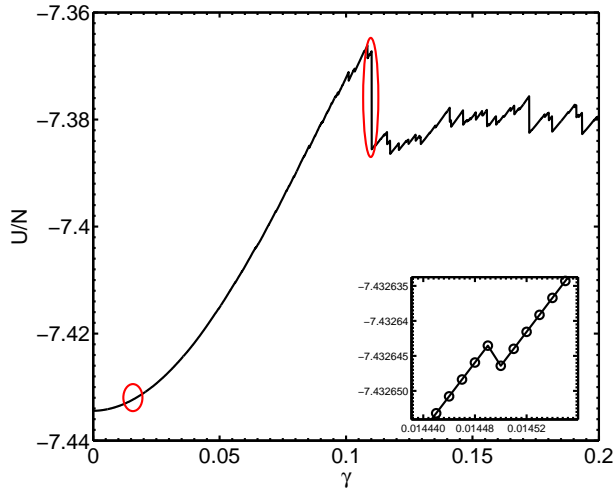}
\caption{(Color Online): Typical dependence of potential energy per particle $U/N$  on strain $\gamma$ in the AQS protocol for the 3-D amorphous solid (N=85184). The inset is a blow-up of the first plastic event which occurs here at $\gamma\approx 0.0145$. The displacement field associated with a similar event is shown in Fig. \ref{elementary}, from which the parameters of the Eshelby model are estimated. The first sub-extensive event is emphasized  by the second red ellipse at $\gamma\approx 0.11$. }
\label{enevsgam}
\end{figure}

 When one gradually increases the external strain in this AQS protocol and allows the system to relax to the mechanical
 equilibrium after every strain increase step, one observes a typical stress-strain curve as seen
 in Fig.\ref{sigvsgam} and an energy vs. strain curve as seen in Fig.\ref{enevsgam}.

 The individual plastic event in the reported simulation is encircled with a red ellipse and is blown up in the inset
 of Fig.\ref{enevsgam}. Like all plastic events, also this one
 occurs when a nonzero eigenvalue $\lambda_P$ of $\B H$ approaches zero at some strain value $\gamma_P$. It was proven that this occurs universally through a saddle-node bifurcation such that $\lambda_P$ tends to zero like $\lambda_P\sim \sqrt{\gamma_P-\gamma}$ \cite{12DKP}. For values of the stress which are below the yield stress the plastic instability is seen \cite{ML} as
 a localization of the eigenfunction of $\B H$, denoted as $\Psi_P$ which is associated with the eigenvalue $\lambda_P$, (see Fig. \ref{elementary}).
 At $\gamma=0$ all the eigenfunctions associated with low-lying eigenvalues are delocalized, the eigenfunction $\Psi_P$ localizes as $\gamma\to \gamma_P$ (when $\lambda_P\to 0$). This eigenfunction represents the displacement field associated with the plastic irreversible stress and energy release.
 \subsection{Analysis and identification of the elementary plastic event}
 \label{octupole}

\begin{figure}
\vskip 0.5 cm
\includegraphics[scale = 0.35]{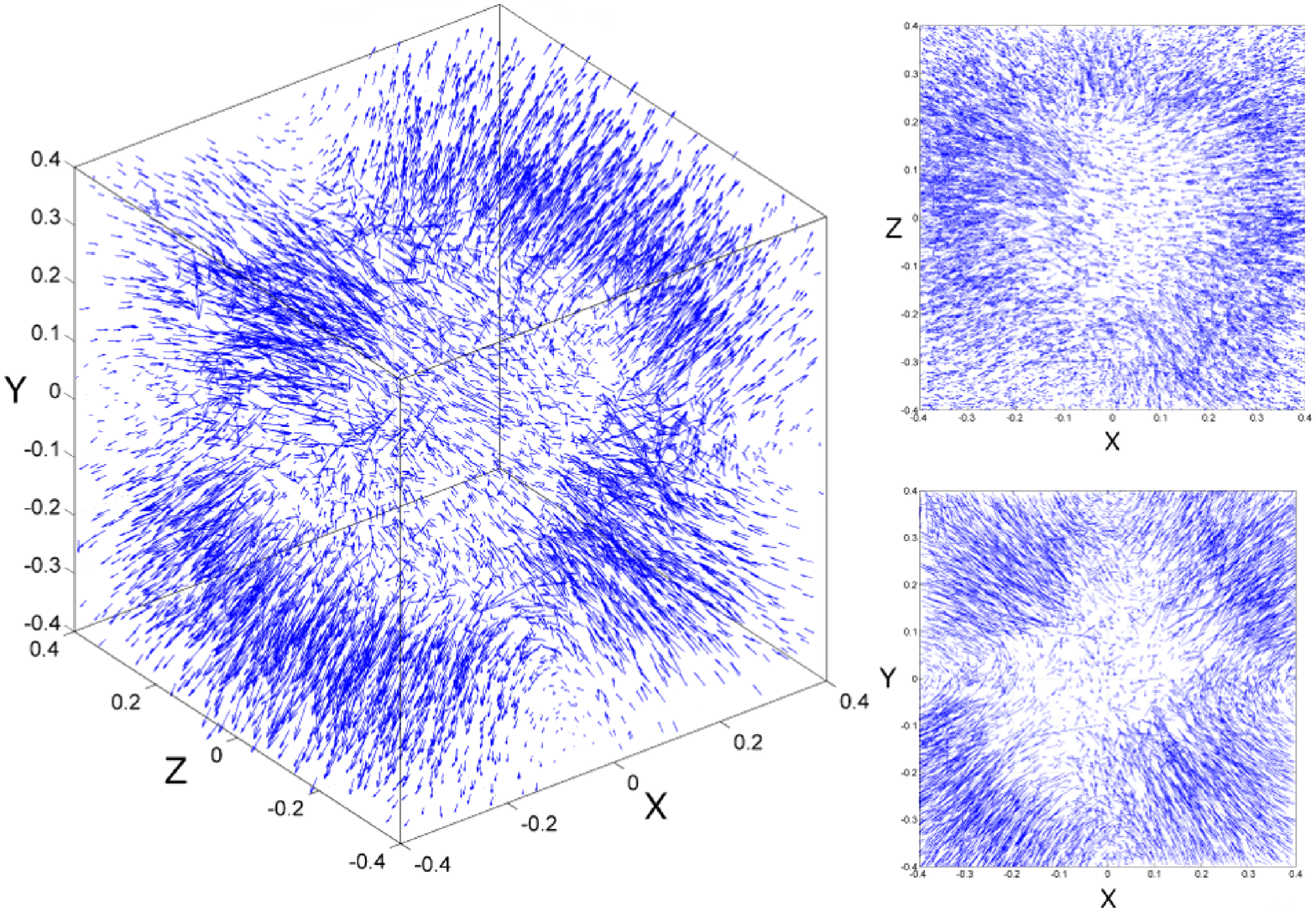}
\includegraphics[scale = 0.35]{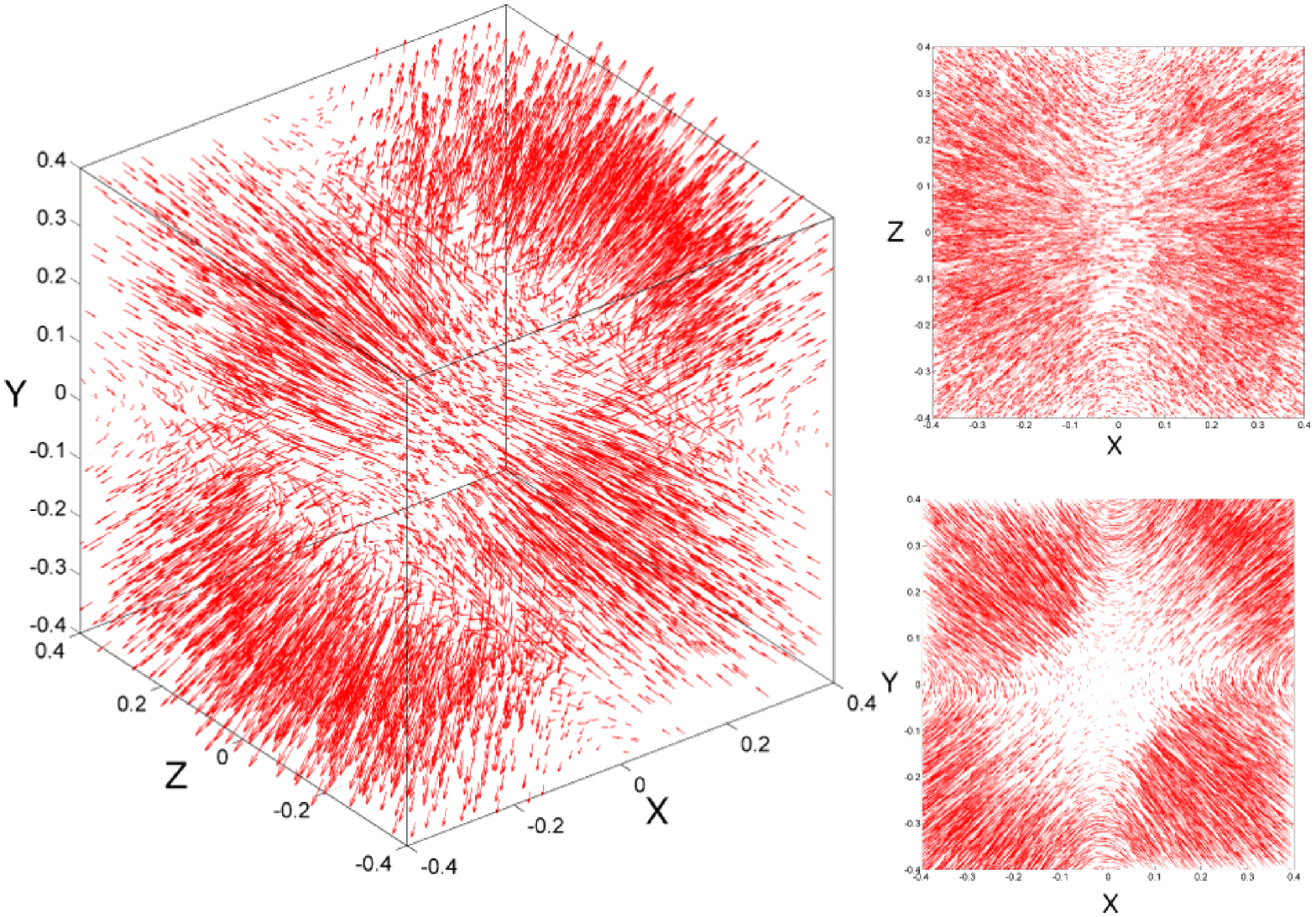}
\caption{(Color Online): Upper panel: the non-affine displacement field associated with an elementary plastic instability similar to the one shown in the inset of Fig. \ref{enevsgam} for N=64000. As mentioned, the local behavior of the non-affine displacement field reflects the localization of the Hessian matrix eigenfunction. To be able to see the event clearly, we show the projections on $X-Z$ and  $X-Y$ planes. For clarity of view we removed a sphere of radius 0.2 from the core of the event. Lower panel: the displacement field of a general 3-dimensional traceless Eshelby inclusion with two nonzero eigenvalues. We used $\epsilon^*=0.04$ and removed a core of radius 0.2.}
\label{elementary}
\end{figure}

 A-priori it is not obvious what is the best Eshelby inclusion that would model faithfully the elementary plastic
 events in 3-D. To shed light on this issue we identified and isolated the elementary plastic events that occur
 in our simulations at low value of $\gamma$.
 In Fig. \ref{elementary} we show this elementary event from two angles of view.
Examining these figures we find that the single event is somewhat similar to the two-dimensional one, with one outgoing direction, one incoming direction, and one neutral direction. In subsect.~\ref{orient} we will substantiate this
finding analytically by showing that this is a minimal energy solution for the three dimensional event under
the conditions of simple shear. Nevertheless, for the sake of generality,
we will offer in the next section a theory in which the Eshelby inclusion has three independent directions.

\section{Theory of elementary plastic events in 3-dimensions}
\label{theory}

Let us consider the spherical Eshelby inclusion which is strained into a general ellipsoid
whose internal strain tensor can be written in general using three orthogonal unit vectors $\hat{\B n} \perp \hat{\B k} \perp \hat{\B m}$ as:
\begin{equation}
 \epsilon^*_{\alpha\beta}  =
\lambda_{n}n_{\alpha}n_{\beta}
+\lambda_{k}k_{\alpha}k_{\beta}
+\lambda_{m}m_{\alpha}m_{\beta} \ ,
\label{genten}
\end{equation}
where the eigenvalues $\lambda_i$ are related to the determinant of the deformation tensor $\B F$  \cite{EranLN} such that
\begin{equation}
\det \B F =(1+\lambda_n)(1+\lambda_k)(1+\lambda_m).
\end{equation}
Volume conservation under the conditions of pure external stress implies $\det \B F=1$.
In a linear theory this volume conservation constraint translates (to first order in the eigenvalues) to the demand
 \begin{equation}
 \lambda_n+\lambda_k+\lambda_m=0  \ .
 \end{equation}
In addition, for an orthogonal triad of unit vectors $\hat n$, $\hat k$ and $\hat m$ we have the relation
\begin{eqnarray}
n_{\alpha}n_{\beta} + k_{\alpha}k_{\beta} + m_{\alpha}m_{\beta} = \delta_{\alpha\beta} \ .
\end{eqnarray}
Using the last two relations, Eq.~(\ref{genten}) for the internal strain tensor of the Eshelby inclusion is reduced to a form:
\begin{eqnarray}
\epsilon^{*}_{\alpha \beta}\!\! =\!\!(2\lambda_n\!\!+\!\!\lambda_k)n_{\alpha}n_{\beta}\!+ \! (2\lambda_k\!+\!\lambda_n)k_{\alpha}k_{\beta}\!-\!(\lambda_n\!+\!\lambda_k)\delta_{\alpha\beta}.
\label{Eshstrain}
\end{eqnarray}

\subsection{Solutions of  Navier-Lam\'e equations in 3D and  computation of the non-affine displacement field.}
\label{disp}
The next step in the analysis of the fundamental plastic event is a solution for the displacement field, the strain field and the stress field associated with the Eshelby inclusion whose strain tensor is given in general form by Eq.~(\ref{Eshstrain}).

The displacement field, denoted as  $u^{c}(\B X)$, should satisfy
 Navier-Lam\'e equations subject to appropriate boundary conditions at the inclusion
boundary. Besides, this displacement field should decay to zero at infinity. In 3D, the Navier-Lam\'e equations with zero
body forces can be written as
\begin{eqnarray}
\frac{1}{1-2\nu}\frac{\partial^2 u_\gamma^c}{\partial X_\alpha \partial X_\gamma}+\frac{\partial^2 u_\alpha^c}{\partial X_\beta \partial X_\beta}=0,
\label{NL}
\end{eqnarray}
where $\B X$ is the position vector centered on the inclusion and
$R \equiv|\B X|$. All solutions of  Eq.~(\ref{NL}) also obey the higher order bi-harmonic equation \cite{Landau}
\begin{eqnarray}\label{BL}
\frac{\partial^4 u^{c}_{\alpha}}{\partial X_{\beta} \partial X_{\beta} \partial X_{\psi} \partial X_{\psi}} = \nabla^2 \nabla^2 u^c_{\alpha} = 0 .
\end{eqnarray}
Now we construct from the  radial solutions of the bi-Laplacian equation Eq.~(\ref{BL}) the derivatives which also satisfy Eq.~(\ref{NL}). Note that Eq.~(\ref{BL}) is only a necessary (but not sufficient) condition for the solutions and Eq.~(\ref{NL}) still needs to be satisfied. Solutions of Eq. (\ref{NL})
which are linear in the eigenstrain and decay to zero as $R\to \infty$
are constructed from linear combinations of the
centrally symmetric solutions of Eq. (\ref{BL}, i.e. $R^2$, $R$, $1$
and $1/R$:
\begin{eqnarray}
u_{\alpha}^{c}&=&
A\epsilon_{\alpha\eta}^{*}\frac{\partial}{\partial X_{\eta}}\left(\frac{1}{R}\right)
+B\epsilon_{\eta\lambda}^{*}\frac{\partial^{3}}{\partial X_{\alpha}\partial X_{\eta}\partial X_{\lambda}}\left(\frac{1}{R}\right)\nonumber\\
&+& C\epsilon_{\eta\lambda}^{*}\frac{\partial^{3}}{\partial X_{\alpha}\partial X_{\eta}\partial X_{\lambda}}R \ .
\label{lincom}
\end{eqnarray}
The calculation of the coefficients is presented in Appendix \ref{disfield}. The final result is
\begin{eqnarray}
u_{\alpha}^{c} & = &
\frac{1}{3(1-\nu)}\frac{a^{3}}{R^{3}}
 \left[(1-2\nu)+\frac{3}{5}\frac{a^{2}}{R^{2}}\right]
\Bigl( (2\lambda_n-\lambda_k) \hat{n}_{\alpha}( \B {\hat{n}} \cdot \B{X})  \nonumber \\
&& \qquad \qquad + (\lambda_n+2\lambda_k) \hat{k}_{\alpha} (\B{\hat{k}} \cdot \B{X}) - (\lambda_n+\lambda_k)X_{\alpha} \Bigr)\nonumber \\
 & + & \frac{\epsilon^{*}}{2(1-\nu)}\frac{a^{3}}{R^{5}}
 \left[1-\frac{a^{2}}{R^{2}}\right]
\Bigl( (2\lambda_n-\lambda_k) ( \B {\hat{n}} \cdot \B{X})^{2}  \nonumber \\
&&  \qquad \qquad +  (\lambda_n-2\lambda_k)(\B{\hat{k}} \cdot \B{X})^2  - (\lambda_n+\lambda_k)\Bigr) X_\alpha
\label{Uc}
\end{eqnarray}
where a hat over a vector represents a unit vector in the direction of that vector.

\subsection{The strain and stress fields induced by the 3D Eshelby inclusion}
\label{strstr}
The expression for the displacement field Eq.~(\ref{Uc}) is the starting point to derive expressions for the constrained strain and stress, which are necessary for the total energy calculation.
We use the identities
\begin{eqnarray}
\frac{\partial f(R)}{\partial X_\beta} & =&f'(R)\frac{X_\beta}{R}\\
\frac{\partial(\hat{\B n}\cdot\B X)}{\partial X_\beta} & =&\hat{n}_\beta
\label{iden}
\end{eqnarray}

Then, we get
\begin{widetext}
\begin{align}
 \frac {\partial u^c_{\alpha} }{\partial X_{\beta}} = &
 \frac{1}{3(1-\nu)}
 \Biggl[-3 \frac{a^3}{R^5}(1-2\nu)-3\frac{a^5}{R^7} \Biggr]\left((2\lambda_n+\lambda_k)\hat{n}_\alpha(\hat{\B n}\cdot  \B{X}) +(\lambda_n+2\lambda_k)\hat{k}_\alpha(\hat{\B k}\cdot  \B{X})-(\lambda_n+\lambda_k)X_{\alpha}   \right)
 X_{\beta}   \nonumber \\
 +& \frac{1}{3(1-\nu)} \left[(1-2\nu)\frac{a^3}{R^3}+\frac{3}{5}\frac{a^5}{R^5} \right]
 \left((2\lambda_n+\lambda_k)\hat{n}_\alpha\hat{n}_\beta +(\lambda_n+2\lambda_k)\hat{k}_\alpha\hat{k}_\beta-
 (\lambda_n+\lambda_k)\delta_{\alpha \beta} \right) \nonumber  \\
 +& \frac{1}{2(1-\nu)}
  \left[-3 \frac{a^3}{R^5}+5\frac{a^5}{R^7}\right]  \left( (2\lambda_n+\lambda_k)\frac{ (\hat{\B n}\cdot  \B{X}) ^2}{R^2} + (\lambda_n+2\lambda_k) \frac{(\hat{\B k}\cdot  \B{X}) ^2 }{R^2}-(\lambda_n+\lambda_k) \right) X_{\alpha}   X_{\beta}   \nonumber \\
+& \frac{1}{2(1-\nu)}
  \left[ \frac{a^3}{R^3}-\frac{a^5}{R^5}\right] \Biggl(
(2\lambda_n+\lambda_k) \left[\frac{2(\hat{\B n}\cdot  \B{X})\hat{n}_\beta}{R^2}- \frac{2(\hat{\B n}\cdot  \B{X})^2 X_{\beta}  }{R^4} \right]+
(\lambda_n+2\lambda_k) \left[\frac{2(\hat{\B k}\cdot  \B{X})\hat{k}_\beta}{R^2}- \frac{2(\hat{\B k}\cdot  \B{X})^2 X_{\beta}  }{R^4} \right]
												\Biggr) X_{\alpha}  \nonumber \\
+& \frac{1}{2(1-\nu)}
  \left[ \frac{a^3}{R^3}-\frac{a^5}{R^5}\right]  \left( (2\lambda_n+\lambda_k)\frac{ (\hat{\B n}\cdot  \B{X}) ^2}{R^2} + (\lambda_n+2\lambda_k) \frac{(\hat{\B k}\cdot  \B{X}) ^2 }{R^2}-(\lambda_n+\lambda_k) \right) \delta_{\alpha \beta} \ .
\end{align}
In the linear approximation $\epsilon_{\alpha\beta}^{c} =\frac{1}{2} \left(\frac{\partial u_{\alpha}^{c}}{\partial X_{\beta}} + \frac{\partial u_{\beta}^{c}}{\partial X_{\alpha}} \right)$:
\begin{align}
\epsilon_{\alpha\beta}^{c}( \B{X})=
\frac{1}{2(1-\nu)}\frac{a^{3}}{R^{3}} & \left[-(1-2\nu)-\frac{a^{2}}{R^{2}}\right]
\Biggl(
	 (2 \lambda_n+\lambda_k)
	 \frac{(\hat{\B n}\cdot \B{X})}{R}  \left(\frac{\hat{n}_\alpha X_{\beta}  }{R}
	+\frac{\hat{n}_\beta X_{\alpha}  }{R}\right) \nonumber \\
	 & \qquad + ( \lambda_n+2\lambda_k)
	\frac{(\hat{\B k}\cdot \B{X})}{R} \left(\frac{\hat{k}_\alpha X_{\beta}  }{R}
	+\frac{\hat{k}_\beta X_{\alpha}  }{R}\right)
	-2(\lambda_n+\lambda_k)
	\frac{X_\alpha X_\beta}{R^2}
\Biggr) \nonumber \\
+\frac{1}{3(1-\nu)}  \frac{a^{3}}{R^{3}} & \left[(1-2\nu)+\frac{3}{5}\frac{a^{2}}{R^{2}}\right]
 \left( (2\lambda_n+\lambda_k)\hat{n}_\alpha\hat{n}_\beta +(\lambda_n+2\lambda_k)\hat{k}_\alpha\hat{k}_\beta-
 (\lambda_n+\lambda_k)\delta_{\alpha \beta} \right)   \\
+\frac{\epsilon^{*}}{2(1-\nu)}  \frac{a^{3}}{R^{3}} & \left[-5+7\frac{a^{2}}{R^{2}}\right]
\left(  (2 \lambda_n+\lambda_k) \frac{(\hat{\B n}\cdot \B{X})^{2}}{R^{2}}
+ ( \lambda_n+2\lambda_k) \frac{(\hat{\B k}\cdot \B{X})^{2}}{R^{2}}
-2(\lambda_n+\lambda_k)\right)
 \frac{X_{\alpha}  X_{\beta}  }{R^{2}} \nonumber \\
+\frac{\epsilon^{*}}{2(1-\nu)}\frac{a^{3}}{R^{3}} & \left[1-\frac{a^{2}}{R^{2}} \right]
\Biggl(
	 (2 \lambda_n+\lambda_k)
	 \left[\frac{(\hat{\B n}\cdot \B{X})}{R}\left(\frac{\hat{n}_\beta X_{\alpha}  }{R} +\frac{\hat{n}_\alpha X_{\beta}  }{R}\right)\right]
	+ ( \lambda_n+2\lambda_k)
	\left[\frac{(\hat{\B k}\cdot \B{X})}{R}\left(\frac{\hat{k}_\beta X_{\alpha}  }{R} +\frac{\hat{k}_\alpha X_{\beta}  }{R}\right)\right]
	 \Biggr) \nonumber \\
+\frac{\epsilon^{*}}{2(1-\nu)}\frac{a^{3}}{R^{3}} & \left[1-\frac{a^{2}}{R^{2}}\right]
\left(  (2 \lambda_n+\lambda_k)
		\frac{(\hat{\B n}\cdot \B{X})^{2}}{R^{2}}
	+ ( \lambda_n+2\lambda_k)
	 	\frac{(\hat{\B k}\cdot \B{X})^{2}}{R^{2}}
	 - ( \lambda_n+\lambda_k) \right)\delta_{\alpha\beta} \ . \nonumber
\end{align}
 \\
 After that, one can use the constrained strain to calculate the constrained
stress with the help of Hooke's law:
\begin{equation}
\sigma_{\alpha\beta}^{c}=2\mu\epsilon_{\alpha\beta}^{c}+\lambda\epsilon_{\eta\eta}^{c}\delta_{\alpha\beta}
\end{equation}
Now we have the expressions for $\epsilon^c_{\alpha \beta}$, $\sigma^c_{\alpha \beta}$ and $u^c_{\alpha \beta}$
which will be used for the calculation of the elastic energy.
\end{widetext}

\begin{figure}[h]
\vskip 0.5 cm
\includegraphics[scale = 1.9]{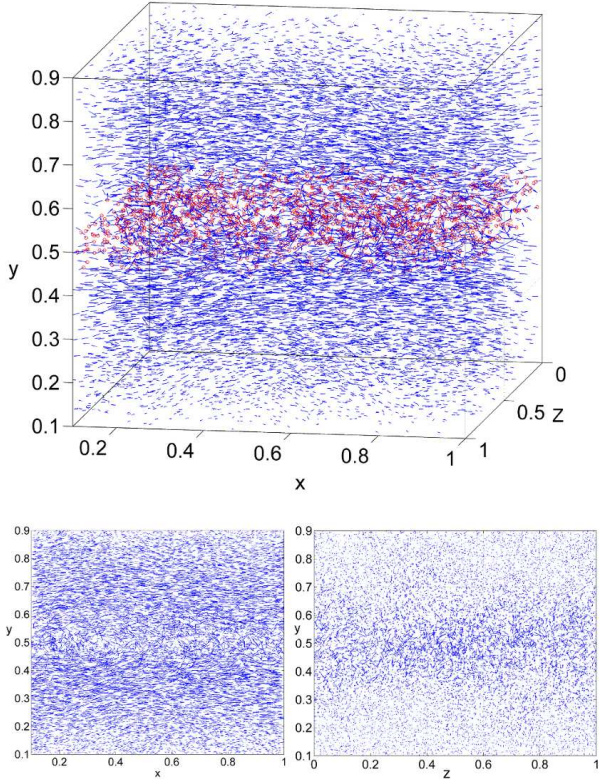}
\caption{(Color Online): The non-affine displacement associated with the system spanning event, i.e. a bigger drop in energy which is encircled in red in Fig. \ref{enevsgam} around $\gamma\approx 0.11$.  One sees that the big drop in energy is associated with the localization of strain in a plane parallel to $\hat{\B{x}}-\hat{\B{z}}$. The red circles at the upper figure were added to guide the eye through the 3D plot. We added an $X-Y$ and a $Z-Y$ projections, in which the nature of the band can be seen more clearly.  On the $X-Y$ plane one observes that the lower part goes leftwards while the upper part moves rightward; these parts are separated by the shear band. On the $Z-Y$ plane the localization is even more clear as one can observe the significant difference in magnitude between the displacement in $Z$ direction (perpendicular to the strain direction) inside and outside the band.}
\label{fullband}
\end{figure}

\section{Numerical Experiments: the shear localizing system-spanning Event}
\label{numexloc}

At this point we return to our numerical simulations in order to ascertain how the elementary events
are organized in space at the appearance of the first shear localizing event.
In the next section we will provide the theory necessary to rationalize the results seen in this section.

In Fig. \ref{fullband} one can observe that the shear localization event is system spanning. For the 2-D case it was shown that the associated energy drop scales with the system size \cite{09LP}.

\section{Theory of system-spanning event}
\label{energysect}

In this section we provide a theory that rationalizes the appearance of shear localization instability as presented
in Fig.~\ref{fullband}. To that aim we start with calculation of the energy associated with a random collection of
$\C N$ inclusions. Then we demonstrate that above some critical value of the external strain the energy approaches a minimum in a  specific configuration when all the inclusions are
organized in a 2-dimensional lattice on the plane (Fig.~\ref{fullband}).

\subsection{The energy associated with $\C N$ inclusions for an external shear strain $\gamma$}

We turn now to calculation of the energy associated with the simultaneous
emergence  of $\C N$ Eshelby inclusions.  We write the energy of $\C N$ inclusions situated  anywhere in the volume:
\begin {equation}
E=\frac{1}{2}\sum_{i=1}^{\C N} \int_{v_{0}^{(i)}} \sigma^{(i)}_{\alpha \beta} \epsilon^{(i)}_{\alpha \beta} dV
+\frac{1}{2} \int_{V- \sum_{i=1}^{\C N}v_{0}^{(i)}} \sigma^{(m)}_{\alpha \beta} \epsilon^{(m)}_{\alpha \beta} dV.
\end{equation}
Here $v_{0}^{(i)}$ is the volume occupied by the $i$th inclusion, the first term represents the contribution from the inclusions themselves and the second part is the matrix deformation contribution, represented by the superscript $m$.
Introducing the notation $A_{\alpha,\beta}\equiv \partial A_\alpha/\partial X_\beta$,
using $\epsilon_{\alpha \beta}=\tfrac{1}{2} (u_{\alpha,\beta}+ u_{\beta,\alpha})$  and  the symmetry of the stress tensor, we obtain:

\begin {equation}
E=\frac{1}{2}\sum_{i=1}^{\C N} \int_{v_0^{(i)}} \sigma^{(i)}_{\alpha \beta} u^{(i)}_{\alpha, \beta} dV
+ \frac{1}{2} \int_{V- \sum_{i=1}^{\C N}v_0^{(i)}}\sigma^{(m)}_{\alpha \beta} u^{(m)}_{\alpha, \beta} dV \ .
\label{EofN}
\end{equation}
This expression is analyzed in Appendix \ref{energy}. The upshot of this analysis is that we need to consider
four terms in the energy, $E=E_{\rm mat}+E_{\infty}+E_{\rm Esh}+E_{\rm int}$, where each of these terms reads
\begin{eqnarray}
E_{\rm mat}&= & \frac{1}{2}\sigma_{\alpha\beta}^{(\infty)}\epsilon_{\alpha\beta}^{(\infty)}V\ ,\\
E_\infty&=&-\frac{1}{2}\sigma_{\alpha\beta}^{(\infty)}\left(\sum_{i=1}^{N}\epsilon_{\alpha\beta}^{(*,i)}v_{0}^{(i)}\right)\ ,\\
E_{\rm Esh} &=& \frac{1}{2}\sum_{i=1}^{N}(\sigma_{\alpha\beta}^{(*,i)}-\sigma_{\alpha\beta}^{(c,i)})
\epsilon_{\alpha\beta}^{(*,i)}v_{0}^{(i)}\ ,\\
E_{\rm int}&=&-\frac{1}{2}\sum_{i=1}^{N}\epsilon_{\alpha\beta}^{(*,i)}v_{0}^{(i)}\left(\sum_{j\neq i}\sigma_{\alpha\beta}^{(c,j)}(\B X_{ij})\right) \ ,
\end{eqnarray}
where $\B X_{ij}$ is the radius vector separating the $i$'th and $j$'th inclusions.

Our aim is to find the configuration of inclusions that minimizes the total energy.
The first term $E_{mat}$ is the matrix energy and does not depend on the inclusions. The $E_{Esh}$ term, is the sum of the "self energy" of $\C{N}$  inclusions:
\begin{align}
E_{\rm Esh}&=\frac{1}{2} v_{0} \sum_{i=1}^{N} \epsilon^{(*,i)}_{\alpha \beta} \left( \sigma^{(*,i)}_{\alpha \beta}-\sigma^{(c,i)}_{\alpha \beta} \right) \nonumber \\
		&=\frac{1}{2} v_{0} \sum_{i=1}^{N} \epsilon^{(*,i)}_{\alpha \beta} C_{\alpha \beta \gamma \delta} \left( \epsilon^{(*,i)}_{\gamma \delta}- \epsilon^{(c,i)}_{\gamma \delta} \right) \nonumber \\
		&=\frac{1}{2} v_{0} \sum_{i=1}^{N} \epsilon^{(*,i)}_{\alpha \beta} C_{\alpha \beta \gamma \delta} \left( \epsilon^{(*,i)}_{\gamma \delta}- \frac{2(4-5 \nu)}{15(1-\nu)}\epsilon^{(*,i)}_{\gamma \delta} \right)   \nonumber \\
		&=\frac{1}{2} v_{0} \sum_{i=1}^{N} \epsilon^{(*,i)}_{\alpha \beta} C_{\alpha \beta \gamma \delta}\epsilon^{(*,i)}_{\gamma \delta}  \frac{(7-5 \nu)}{15(1-\nu)} \ .
\end{align}
Plugging  $C_{\alpha \beta \gamma \delta} \epsilon^{(*,i)}_{\gamma \delta} \equiv
			\mu(\delta_{\alpha \gamma}\delta_{\beta \delta}+\delta_{\alpha \delta}\delta_{\beta \gamma})\epsilon^{(*,i)}_{\gamma \delta}
			+\lambda\delta_{\alpha \beta}\delta_{\gamma \delta}\epsilon^{(*,i)}_{\gamma \delta}$
and using Young modulus $\C E$ instead of shear modulus $\mu$ one obtains:
\begin{align}
E_{\rm Esh} &=\frac{1}{2} v_{0}\mathcal{E}  \frac{(7-5 \nu)}{15(1-\nu^2)} \sum_{i=1}^{N} \epsilon^{(*,i)}_{\alpha \beta} \epsilon^{(*,i)}_{\beta \alpha}.
\end{align}
Replacing $\epsilon^{(*,i)}_{\alpha \beta}$ with its expression from Eq. (\ref{Eshstrain}) we have:
\begin{align}
E_{\rm Esh} &=\frac{1}{2} v_{0}\mathcal{E}  \frac{(7-5 \nu)}{15(1-\nu^2)} \sum_{i=1}^{N}(\lambda_n^2+\lambda_k^2+(\lambda_n+\lambda_k)^2).
\label{Eesh}
\end{align}
Here we assume that all the inclusions have the same eigenvalues.

Next, we note that the only
term that contains the external strain is $E_\infty$, so for large enough external strain we must minimize this
term separately. For this term we have $\epsilon_{xy}^{\infty}=\tfrac{\gamma}{2}\:\:,\:\:\epsilon_{xz}^{\infty}=\epsilon_{zy}^{\infty}=\epsilon_{ii}^{\infty}=0$,
resulting in $\sigma_{xy}^{\infty}=\mu\gamma$. Using Eq. (\ref{genten}) we find for $\C N$ identical inclusions
\begin{eqnarray}
E_{\infty}=
-v_0 \mu \frac{\gamma}{2} \sum_{i=1}^N & \biggl( \lambda_n n^{(i)}_{x}n^{(i)}_{y} +\lambda_k k^{(i)}_{x}k^{(i)}_{y} \nonumber \\
&-(\lambda_n+\lambda_k) m^{(i)}_{x}m^{(i)}_{y} \biggr)
\end{eqnarray}

Considering the most general form of an orthogonal triad $\hat{n},\hat{k},\hat{m}$ and
 writing $\hat{n},\hat{k}$ in polar coordinates (with $\psi$ the rotation angle of $\hat{k}$ around the $\hat z,\hat{n}$ plain), one obtains:
\begin{eqnarray}
\hat{n}_x&= &\sin(\theta)\cos(\phi)\nonumber\\
\hat{n}_y&= &\sin(\theta)\sin(\phi)\nonumber\\
\hat{n}_z&= &\cos(\theta) \nonumber\\
 \nonumber\\
\hat{k}_x&= &-\cos(\theta)\cos(\phi)\cos(\psi)-\sin(\phi)\sin(\psi) \nonumber \\
\hat{k}_y&= &-\cos(\theta)\sin(\phi)\cos(\psi)+\cos(\phi)\sin(\psi)\nonumber\\
\hat{k}_z&= &\sin(\theta)\cos(\psi)\nonumber \\
 \nonumber\\
\hat{m}_x&= &-\cos(\theta)\cos(\phi)\sin(\psi)+\sin(\phi)\cos(\psi) \nonumber \\
\hat{m}_y&= &-\cos(\theta)\sin(\phi)\sin(\psi)-\cos(\phi)\cos(\psi) \nonumber\\
\hat{m}_z&=  &\sin(\theta)\sin(\psi) \ .
\end{eqnarray}
Writing $E_{\infty}$ in polar coordinates, we have:
\begin{widetext}
\begin{eqnarray}
&&E_{\infty}^{(i)}=  -v_{0}\mu\frac{\gamma}{2}  \left[\biggl( \lambda_n n^{(i)}_{x}n^{(i)}_{y} +\lambda_k k^{(i)}_{x}k^{(i)}_{y}-(\lambda_n+\lambda_k) m^{(i)}_{x}m^{(i)}_{y} \biggr)\right] \nonumber \\
&&= -v_{0}\mu \frac{\gamma}{4}\Big[\lambda_n
	 \bigl( \sin(2\phi) (\sin^2(\theta)-\cos^2(\theta)\sin^2(\psi)+\cos(\psi)	 )
	 -\cos(2\phi) \cos(\theta) \sin(2\psi )  \bigr) \nonumber \\
&& \qquad \quad + \lambda_k
	\big(\sin(2\phi ) \cos(2\psi )( \cos^2(\theta)+1 )
	+2\cos(2\phi)\cos(\theta)\sin(2\psi ) \big) \Big] \ .
 \label{Einfty}
\end{eqnarray}
\end{widetext}

\subsection{The orientation of each elementary inclusion at high enough external strain}
\label{orient}

We now have an expression for $E_{\infty}=E_\infty(\theta,\phi,\psi,\lambda_n,\lambda_k)$. At this point we need
to find the minimal energy for each inclusion under the constraint of constant self-energy $E_{\rm Esh}$ .
This way we'll determine the shape and the orientation of the most energetically favorable inclusion under given conditions of the pure shear.   Setting $E_{Esh}$ from Eq. (\ref{Eesh}) to a constant value will provide a relation between $\lambda_n$ and $\lambda_k$ in the following form:
\begin{align}
 \lambda_k&=\tfrac{1}{2} (-\lambda_n \pm \sqrt{-3 \lambda_n^2+4\tilde{E}}) \ ,\label{nice} \\
 \tilde E&=\frac{E_{Esh}}{v_{0}\mathcal{E} } \frac{15(1-\nu^2)}{(7-5 \nu)}
 \end{align}
Using this, we obtain an expression for $E_\infty^{(i)}=E_\infty^{(i)}(\theta,\phi,\psi,\lambda_n)$.
The minimum of this expression is attained at $( \theta, \phi, \psi,\lambda_n)=(\tfrac{\pi}{2},\tfrac{\pi}{4},\tfrac{\pi}{2},\sqrt{\tilde{E}})$.
 This implies that the $\hat{\B n},\hat{\B k}$ plane must coincide with the $\hat {\B x},\hat {\B y}$ plane and  the eigenvalue associated with $\hat{\B z}$ axis ($\hat{m}$) vanishes. To see this one just needs to
 plug the value $\lambda_n=\sqrt{\tilde E}$ in Eq. (\ref{nice}) to realize that $\lambda_k=-\lambda_n$. In other words, the displacement field associated with the
 minimum energy solution for each inclusion has the same orientation such that ${\hat n}_x={\hat n}_y= \frac{1}{\sqrt{2}}$, and that the $\hat{\B z}$ axis turns out to be ``neutral".

 Having this result, we can now simplify the expression (\ref{Eshstrain} ) for the strain tensor of the single inclusion:
\begin{equation}
\epsilon^*_{\alpha\beta} = \epsilon^{*} n_{\alpha}n_{\beta} - \epsilon^{*} k_{\alpha}k_{\beta}
\label{Eshstrain2}
\end{equation}
where $\epsilon^*$ is  the only remaining free parameter.\\

\subsection{The minimal energy configuration of $\C N$ inclusions}

Next we need to analyze the term denoted as $E_{int}$. This is done in Appendix \ref{Eint} and leads to the following final expression:
\begin{eqnarray}
&&E_{\rm int}= \displaystyle
-\frac{1}{2} \sum_{<i,j>} E_{int}^{ij}  ; \nonumber \\
&&E_{int}^{ij}  =
-v_{0}2\mathcal{E}\frac{a^{3}}{(R_{ij})^{3}}\frac{(\epsilon^{*})^2}{(1-\nu^2)}\times \nonumber\\
&&\Biggl( 4\left[\nu-\frac{a^{2}}{(R_{ij})^{2}}\right]
 \Bigl \{ (\B{\hat{n}}\cdot\B{\hat{r}}^{(ij)})^2
   +(\B{\hat{k}}\cdot\B{\hat{r}}^{(ij)})^2 \Bigr \}\nonumber \\
&&+ \biggl[\frac{2(1-2\nu)}{3}-\frac{2}{5}\frac{a^{2}}{(R_{ij})^{2}}\biggr]
 \Big \{ 2 \Bigr \}
 \nonumber \\
&&+ \left[-5+7\frac{a^{2}}{(R_{ij})^{2}}\right]
\Bigl \{\Bigl( (\B{\hat{n}}\cdot\B{\hat{r}}^{(ij)})^{2}
  -(\B{\hat{k}}\cdot\B{\hat{r}}^{(ij)})^{2} \Bigr)^2
 \Bigr \}  \Biggr) \ .\nonumber \\
\end{eqnarray}
As proved in the Appendix, this expression has a minimum at  $(\hat{\B n} \cdot \hat{\B r}^{(ij)})= (\hat{\B k} \cdot \hat{\B r}^{(ij)})$, that gives the minimal energy for  $(\hat{\B n} \cdot \hat{\B r}^{(ij)})= (\hat{\B k} \cdot \hat{\B r}^{(ij)})=\tfrac{1}{\sqrt{2}}$. It means that the minimal energy configuration for each pair of inclusions is  attained when they are both aligned along a line parallel to the strain direction. Apart from that,  placement of the inclusion on a line parallel to the $\hat{\B x}-\hat{\B z}$ plane is still energetically favorable over any other pairwise configuration. This implies that when  $\C N$ inclusions are placed on a plane parallel to the $\hat{\B x}-\hat{\B z}$ , the energy attains the (local) minimum. Moreover, disposition of  these inclusions on such a plane will form the desired ``shear plane'', that will separate the upper and lower parts of the matrix moving in opposite directions.
Having that in mind, we can also determine the planar structure of the inclusions. Assuming that all inclusions are on the plane parallel to $\hat{\B x}-\hat{\B z}$, we can now write $E_{\rm int}^{(ij)}$ in polar coordinates, where the $\theta=0$ axis is perpendicular both to $\hat{\B n}$ and $\hat{\B k}$:

\begin{align}
 E_{\rm int}= &\displaystyle
-\frac{1}{2} \sum_{<i,j>} E_{int}^{ij}(R_{ij},\theta)  \\
  E_{int}^{ij}(R_{ij},\theta) & =
-8 v_{0}\mathcal{E} \frac{a^{3}}{(R_{ij})^{3}}\frac{(\epsilon^{*})^2}{(1-\nu^2)} \: \times \nonumber \\
\quad \Biggl(
 \bigg[\nu-&\frac{a^{2}}{(R_{ij})^{2}}\bigg]
\sin^2(\theta)
   + \biggl[\frac{(1-2\nu)}{3}-\frac{1}{5}\frac{a^{2}}{(R_{ij})^{2}}\biggr]
   \Biggr)\nonumber
\end{align}
To find the minimal energy configuration, we calculate the $E_{\rm int}$ term for two general lattices: a triangular lattice and a rectangular lattice (Fig \ref{lattices}).

 \begin{widetext}

\begin{figure}[h]
\includegraphics[scale = 0.3]{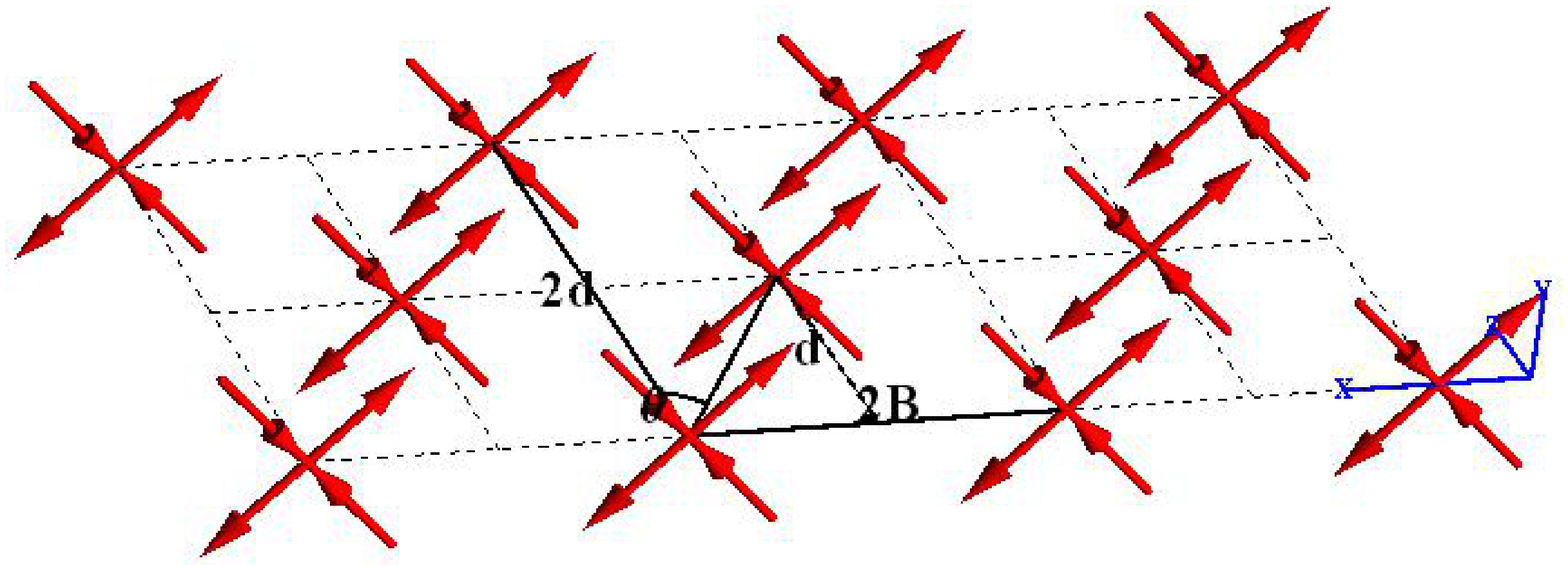}
\includegraphics[scale = 0.3]{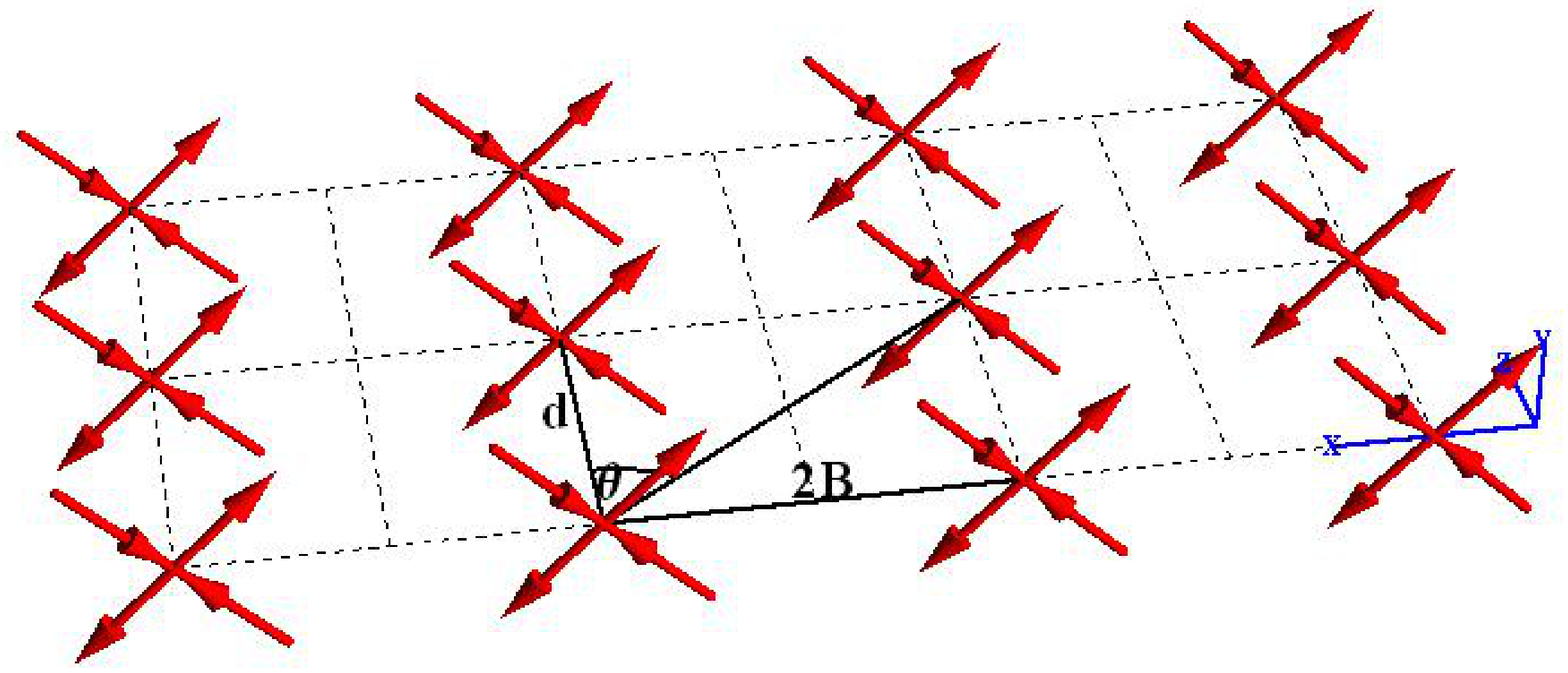}
\caption{(Color Online): Sketch of the two kinds of lattices analyzed . Left: the general non-regular triangular lattice, with 8 neighbors (3 nearest neighbors); right: the equivalent rectangular lattice. Note that the \lq\lq{} lattice points\rq\rq{} in this case have a direction that makes the $45^\circ$ triangular lattice different from the square lattice.  (see fig. \ref{latticeE})}
\label{lattices}
\end{figure}
   \subsubsection{Energy of the triangular lattice with three nearest neighbors}
   Denoting the distances $d$ and $B$ as shown in Fig \ref{lattices}, we can get that the energy per inclusion in this lattice:
 \begin{align}
 \tilde{E}_{int}(R_{ij},\theta)&=2\cdot E({2B,\tfrac{\pi}{2}})+2\cdot E({2d,0}) +2\cdot E({\sqrt{B^2+d^2},\theta}) +2\cdot E({\sqrt{B^2+d^2},-\theta}) \nonumber \\
 \Rightarrow \qquad  &= 2\cdot E^{2B}+2\cdot E^{2d} +4\cdot E^{\theta}.\nonumber \\
 \end{align}
 Calculating each term separately, we find:
  \begin{align}
 E^{2B}&=-v_{0} 2 \mathcal{E} \frac{(\epsilon^{*})^2}{(1-\nu^2)} \frac{a^{3}}{8 B^{3}}
\Biggl(
 4\left[\nu-\frac{a^{2}}{4 B^{2}}\right]
  + \biggl[\frac{4(1-2\nu)}{3}-\frac{4}{5}\frac{a^{2}}{4 B^{2}}\biggr]
  \Biggr)
 \nonumber \\
 E^{2d} &=-v_{0} 2 \mathcal{E} \frac{(\epsilon^{*})^2}{(1-\nu^2)} \frac{a^{3}\tan^3(\theta)}{8B^{3}}
\Biggl(
    \biggl[\frac{4(1-2\nu)}{3}-\frac{4}{5}\frac{a^{2} \tan^2(\theta)}{4 B^{2}}\biggr]
  \Biggr)\nonumber \\
E^{\theta} &=-v_{0} 2 \mathcal{E} \frac{(\epsilon^{*})^2}{(1-\nu^2)} \frac{a^{3}\sin^3(\theta)}{B^3}
\Biggl(
 4\left[\nu-\frac{a^{2} \sin^2(\theta)}{B^2}\right]
\sin^2(\theta)
 +  \biggl[\frac{4(1-2\nu)}{3}-\frac{4}{5}\frac{a^{2} \sin^2(\theta)}{B^2}\biggr]
 \Biggr) \nonumber \\
 \end{align}
 using $E_0=2 v_{0}  \mathcal{E} \frac{ a^2(\epsilon^{*})^2}{(1-\nu^2)}$ and plugging all in:
  \begin{align}
 \tilde{E}_{int}=-E_0\Biggl \{
 			\frac{1}{ B^{3}} \Biggl( &
 			 \nu \left( 1+16 \sin^5(\theta) \right)
 			 + \biggl[\frac{(1-2\nu)}{3} \biggr] \left( 1+\tan^3(\theta)+16 \sin^3(\theta) \right) \nonumber \\
 			 & -  \frac{a^{2}}{B^{2}} \left( \frac{6}{20}+\frac{\tan^5(\theta)}{20}+16\sin^5(\theta)+16 \sin^7(\theta) \right) \quad
 			   \Biggr)		
 			  \Biggr \}
 \end{align}
Taking only leading order terms (without the $\tfrac{a^5}{B^5}$ terms) and replacing  $B$ by the equivalent density term $B^2=\tfrac{\tan(\theta)}{2 \rho}$ (here $\rho$ is the areal density of inclusions on the plane), we have:
  \begin{align}
 \tilde{E}_{int}^{_\triangle}(\rho, \theta)=-E_0\Biggl \{
 			\frac{(2\rho)^{3/2}}{\tan^{3/2}(\theta)} \Biggl(
 			 \nu \left( 1+16 \sin^5(\theta) \right)
 			 + \biggl[\frac{(1-2\nu)}{3} \biggr] \left( 1+\tan^3(\theta)+16 \sin^3(\theta) \right)
 			  \Biggr)		
 			  \Biggr \}
 \end{align}
 Differentiating with respect to  $\theta$ yields:
  \begin{align}
\frac{\partial \tilde{E}_{int}^{_\triangle}}{\partial \theta}=-E_0\Biggl \{
 			\frac{(2\rho)^{3/2}}{\tan^{3/2}(\theta)} \Biggl( &
 			 \nu \left( 80\sin^4(\theta) \cos(\theta) \right) \nonumber \\
 			 &+ \biggl[\frac{(1-2\nu)}{3} \biggr] \left( 3\frac{\tan^3(\theta)}{\cos^2(\theta)}+48 \sin^2(\theta) \cos(\theta) \right)  			   \Biggr)\nonumber \\
 			-\frac{3}{2}\frac{(2\rho)^{3/2}}{\tan^{5/2}(\theta) \cos^2(\theta)}  \Biggl( &
 			 \nu \left( 1+16 \sin^5(\theta) \right)
 			 + \biggl[\frac{(1-2\nu)}{3} \biggr] \left( 1+\tan^3(\theta)+16 \sin^3(\theta) \right)
 			  \Biggr)		
 			  	  \Biggr \}   \nonumber
 \end{align}
  As one can see in Fig~\ref{latticeE}, this energy term has one non-trivial minimum slightly above $\theta \approx \tfrac{\pi}{4}$  (the exact value depends on the Poisson ratio).

\begin{figure}
\includegraphics[scale = 0.25]{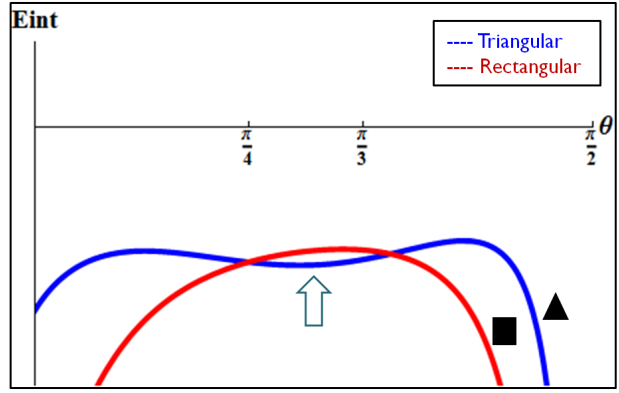}
\caption{The $E_\infty$ energy density for the triangular lattice (blue) and for the rectangular lattice (red). The triangular lattice configuration has a local minimum, whereas the rectangular does not. The minimum is attained for the angle $\theta\approx 50^\circ$}.
\label{latticeE}
\end{figure}
 To check this we also calculated the energy  of a general rectangular lattice, see the next subsection.
 \subsubsection{Energy of the rectangular lattice with three neighbors}
Again, we start with the energy per inclusion:
  \begin{align}
 \tilde{E}_{int}&=2\cdot E(d,0)+2\cdot E(2B,\tfrac{\pi}{2}) +4\cdot E(\sqrt{4B^2+d^2},\theta) \ . \nonumber
\end{align}
Calculating the energy in the same manner as in the case of the triangular lattice (while keeping the same $\rho=2Bd$ relation - see Fig. (\ref{lattices}) ), we obtain the expression in the far-field approximation:
  \begin{align}
 E^{2B}&=-v_{0} 2 \mathcal{E} \frac{(\epsilon^{*})^2}{(1-\nu^2)} \frac{a^{3}}{8 B^{3}}
\Biggl(4 \nu+ \frac{4(1-2\nu)}{3} \Biggr) \ ,
 \nonumber \\
 E^{d} &=-v_{0} 2 \mathcal{E} \frac{(\epsilon^{*})^2}{(1-\nu^2)} \frac{a^{3}\tan^3(\theta)}{8B^{3}}
\Biggl(  \frac{4(1-2\nu)}{3} \Biggr) \ , \nonumber \\
E^{\theta} &=-v_{0} 2 \mathcal{E} \frac{(\epsilon^{*})^2}{(1-\nu^2)} \frac{a^{3}\sin^3(\theta)}{8B^3}
\Biggl( 4\nu \sin^2(\theta) + \frac{4(1-2\nu)}{3} \Biggr)  \ . \nonumber \\
 \end{align}
  using $E_0=2 v_{0}  \mathcal{E} \frac{ a^2(\epsilon^{*})^2}{(1-\nu^2)}$ and plugging all in:
  \begin{align}
 \tilde{E}_{int}=-E_0\Biggl \{
 			\frac{1}{ B^{3}} \Biggl(  \nu \left( 1+2 \sin^5(\theta) \right)
 			 + \biggl[\frac{(1-2\nu)}{3} \biggr] \left( 1+\tan^3(\theta)+2 \sin^3(\theta) \right)  \Biggr)	
 			   \Biggr \}
 \end{align}
 To keep $\rho=1/2Bd$ we now use the relation $B^2=\tfrac{\tan(\theta)}{4 \rho}$ and thus obtain:
  \begin{align}
  \tilde{E}_{int}^{\diamond}(\rho, \theta)=-E_0\Biggl \{
 			\frac{(4 \rho)^{3/2}}{ \tan^{3/2}(\theta)} \Biggl(  \nu \left( 1+2 \sin^5(\theta) \right)
 			 + \biggl[\frac{(1-2\nu)}{3} \biggr] \left( 1+\tan^3(\theta)+2 \sin^3(\theta) \right)  \Biggr)	
 			   \Biggr \}
 \end{align}
This expression has no minima (see \ref{latticeE}). Even if we allow another degree of freedom  - $\phi$ angle which is the rotation of the lattice with respect to $\hat{\B z}$, we will still get no minimum, but will find that the square lattice is a stationary point. The triangular lattice minimum calculated above is very close to the energy of this square lattice solution, but still has a small deviation that makes it favorable.

It implies that for a given areal density $\rho$ the triangular lattice with this specific $\theta$ angle is the energetically favorable configuration.
 Then, we can conclude that for a given  areal density $\rho$ , it is possible to tile the shear plane with inclusions in the energetically favorable manner. Thus, one can create the desired separation between the two parts of the matrix.
 \begin{figure}
\includegraphics[scale = 0.25]{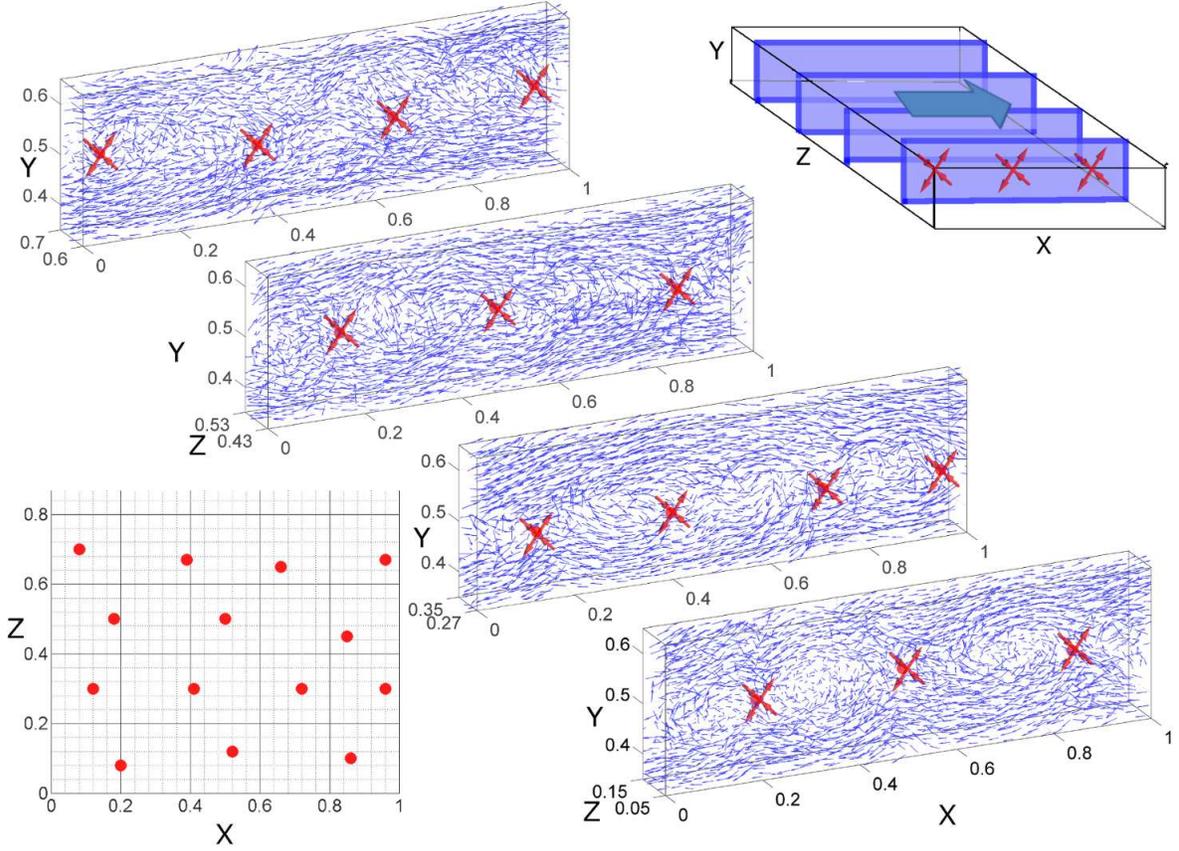}
\caption{Organization of the inclusions on the plane of the shear band. We show thin $X-Y$ cuts taken at different $Z$ sections as illustrated at the top right corner. The crosses represent the suggested centers of the inclusions. At the bottom left corner we present the full picture of the inclusions located in the $X-Z$ shear plane. One can see that the planar configuration of the inclusions corresponds to the triangular lattice as predicted in section \ref{energysect}. The average $\theta$ angle on the reconstructed lattice is approximately $\theta \approx 0.7$ which is somewhat lower than the result predicted at fig. \ref{latticeE}. Most probably this is due to the small system size and the periodic boundary conditions applied.}
\label{bandplane}
\end{figure}
\end{widetext}

\section{Comparison with the numerical simulation - internal structure of the shear band}
\label{compare}
At this point is is important to verify numerically the prediction of the triangular lattice arrangement of the
elementary events in the shear localization plane. To this aim, we need to examine carefully the internal region of
our simulation box and to consider various cross-sections. The results of this analysis are shown in Fig. \ref{bandplane}.
Thin $X-Y$ slices with about 4-5 particles in $Z$ direction are used to visualize the inclusions. When
reconstructed on the $X-Z$ plane the structure shown on the left is very similar to the triangular lattice discussed in the
previous section. Needless to say, one cannot expect to obtain perfect regular arrangement of the inclusions in realistic simulation of the random glass in relatively small cell.

\section{Evaluation of the yield strain}
\label{gamy}
On the basis of the analytic and numeric results presented above, one can accept that the inclusions are indeed arranged in a triangular lattice. Now it is possible to reconsider the total energy of the system of inclusions and to evaluate the critical yield strain. We remind here that the energy is presented as a sum of  four terms, $E=E_{mat}+E_{\infty}+E_{\rm Esh}+E_{int}$. Now we look at the value of each one of these terms for the triangular lattice  with areal density $\rho$ of the inclusions. For analytic simplicity, we will calculate the energy of the triangular lattice with $\theta=\tfrac{\pi}{4}$ which is quite close to the optimal angle of about $50^\circ$. We will then demonstrate that this assumption will not affect the predicted value of the yield strain $\gamma_{_{\rm Y}}$.

Returning to Eq. (\ref{4E}) we need to calculate $E_{\infty}, E_{\rm Esh}$ and $E_{int}$, for the $N$ inclusions triangular lattice. As  $E_{\infty}, E_{\rm Esh}$ does not depend on the spatial placement of the inclusions, for $E_\infty$ one can just use the expression $E_\infty$ from Eq. \ref{Einfty} with the values of the arguments obtained for the minimal energy conditions. Replacing $\mu=\tfrac{\mathcal{E}}{2(1+\nu)}$ to obtain an explicit dependence on the Poisson ratio,  one obtains a final expression for $E_\infty$:
\begin{align}
E_{\infty} &= -\C N v_0  \mathcal{E}  \gamma \epsilon^{*}  \frac{1}{4(1+\nu)}
\end{align}
For evaluation of $E_{\rm Esh}$ we take the expression from Eq. (\ref{Eesh}) and plug in it the expression for $\epsilon^*$ from Eq. \ref{nice}. Thus, we obtain:
\begin{align}
E_{esh} &=\C N v_{0}\mathcal{E} (\epsilon^{*})^2 \frac{(7-5 \nu)}{15(1-\nu^2)}
\end{align}
We define the energy density in narrow (with width of $a$) parallelepiped with area $L^2$. Using $\rho=N/L^2$ and $v_0=\tfrac{4}{3}\pi a^3$ we obtain:
\begin{align}
\frac{E_{\rm Esh}}{L^2 a } &=\tfrac{4}{3} \pi \mathcal{E} (\epsilon^{*})^2 \frac{(7-5 \nu)}{15(1-\nu^2)} \rho a^2 \nonumber \\
\frac{E_{\infty}}{L^2 a} &= -\tfrac{4}{3} \pi \mathcal{E}  \gamma \epsilon^{*}  \frac{1}{4(1+\nu)}  \rho a^2
\end{align}
For $E_{\rm int}$ we will use the result obtained for the 45$^\circ$ triangular lattice:
\begin{align}
&\frac{E_{int}^{\diamond}}{L^2a}=-\tfrac{4}{3} \pi  \mathcal{E} \frac{(\epsilon^{*})^2}{(1-\nu^2)}
 		(2\rho a^2)^{5/2} \Biggl(\frac{(4+\sqrt{2})(\nu-2)}{3\sqrt{2}} 		
 			  \Biggr),		
 			  \nonumber \\
\end{align}
For the sake of simplicity, we denote $\C A=(\frac{(4+\sqrt{2})(\nu-2)}{3\sqrt{2}})$, so the total plastic energy density of the system is expressed as:
\begin{widetext}
\begin{align}
\label{endens}
\frac{E(\rho , \gamma)}{L^2 a } &=
	\frac{E_{esh}+E_{\infty}+E_{lat}}{L^2 a } \nonumber \\
&=\frac{4}{3} \frac{\pi \mathcal{E} (\epsilon^{*})^2 }{(1-\nu^2)}
	\left( \left[ \frac{(7-5 \nu)}{15}-\frac{\gamma (1-\nu)}{4 \epsilon^{*}} \right] \rho a^2
		- \mathcal{A} (2\rho a^2)^{5/2}	\right) \nonumber \\
&=\frac{4}{3} \frac{\pi \mathcal{E} (\epsilon^{*})^2 }{(1-\nu^2)}
	\left( \frac{(7-5 \nu)}{15} \left[1-\frac{15  (1-\nu)}{4 (7-5 \nu) \epsilon^{*}}\gamma \right] \rho a^2
		- \mathcal{A} (2\rho a^2)^{5/2}	\right) \nonumber \\
&=\frac{4}{3} \frac{\pi \mathcal{E} (\epsilon^{*})^2 }{(1-\nu^2)}
	\left( \frac{(7-5 \nu)}{15} \left[1-\frac{\gamma}{\gamma_{_Y}}\right] \rho a^2
		- \mathcal{A}  (2\rho a^2)^{5/2}	\right) \nonumber \\
\end{align}
\end{widetext}
Then, we obtain the following expression for the critical shear strain $\gamma_{_Y}$:
\begin{equation}
\gamma_{_Y} \equiv \frac{4 (7-5 \nu) \epsilon^{*}}{15  (1-\nu)} \ .
\label{gammay}
\end{equation}
Parameters of the elementary plastic events were estimated by fitting the results of numeric simulations to the analytic expressions for the displacement fields of the model Eshelby inclusions. The estimation of $\epsilon^*$ from the single event depicted in Fig.\ref{elementary} yielded the values between 0.04 to 0.06. Plugging in the value of the Poisson ratio $\nu$ (estimated to be around 0.22), we get that $\gamma_{_Y}\approx 0.08-0.12$, which is in good agreement with the actual large instability observed in the numeric simulation and accompanied by formation of the shear band.

We should note at this point that the value of the yield-strain is not sensitive to the nature of the lattice
formed by the inclusions on the shear localization plane. The difference in configurations that could be assumed by
these elementary events contribute only to the order $\rho^{5/2}$ in Eq. \ref{endens}, and not to $\gamma_{_{\rm Y}}$ within the approximation used.

\section{Summary and Conclusions}
\label{conclusions}

The goal of this paper was to suggest an atomistic theory for shear localization in 3-dimensional system. The approach starts from the analytic description of elementary plastic events in amorphous solids; these events are modelled by formation of Eshelby inclusions.  Following the guidelines set in \cite{12DHP,13DHP} shear localization is understood as a system spanning plastic instability in which the elementary plastic events get highly correlated along lines or planes in two and three dimensions respectively. It is shown that when the external strain is too small, the only possible plastic
 events are isolated and not system-spanning. When the external strain exceeds a threshold value the elementary events can appear with the finite density, organized as said on lines or planes.

 In this paper we found that in three dimensions under simple shear this is still the case, with the third allowed eigenvalue being very close to zero. This allows a simple and natural arrangement of the elementary events on a plane such that the global connections between them result in the displacement field that streams in one direction above the shear localization plane and in the other direction below it. It should  be mentioned that for relatively small values of the external strain  one observes also individual localised plastic events with three independent nonzero eigenvalues. For the scenario of collective instability in the case of relatively large strains all observed events were effectively two-dimensional, with the third eigenvalue almost equal to zero.

It should be pointed out here that the alignment of the localization plane in 45$^\circ$ to the principal stress is
particular to the pure shear. For other loading conditions, the angle of the shear bands can lie between $30^\circ$ and $60^\circ$ \cite{13AGPS,13AGPSa}.

 The finding of the predicted triangular lattice in the numerical simulations is yet another supporting evidence to our fundamental assumption of the relation between the small local plastic events and the system-spanning events.  Like in two dimensions, it turns out that for pure shear the expression for $\gamma_{_{\rm Y}}$ contains only the Poisson ratio and the Eshelby eigenstrain. It appears rather reassuring that the numerically estimated values of $\epsilon^*$ and the Poisson ratio provide a fairly accurate prediction for the yield stress as appearing in Eq. (\ref{gammay}). Note that also
 this value will change for different loading conditions as explained in \cite{13AGPS,13AGPSa}.

 The last point that needs further development pertains to the preferred density $\rho$ of elementary events on the
 shear localization plane. Clearly, as long as we employ linear elasticity theory, there is no typical scale
 that can emerge, and one needs to go nonlinear in order to find that length scale. This and related issues
 belong to our future research program.

 \acknowledgements

 This work had been supported in part by the German Israeli Foundation, the Minerva Foundation, Munich Germany,
 and the European Research Council under and ``ideas" STANPAS grant. Useful discussions with George Hentschel
 are gratefully acknowledged. The 3-dimensional parallel code used was developed in part by Ashwin J. to whom we express
 our thanks for sharing it with us.

\appendix
\begin{widetext}
\section{Solutions of  Navier-Lam\`e equation}
\label{disfield}
To arrive from Eq. (\ref{lincom}) to  Eq. (\ref{Uc}), we  first take the Laplacian of $u^c_\alpha$:
\begin{eqnarray}
\frac{\partial^2 u^c_\alpha}{\partial X_\beta \partial X_\beta} & = &
 A\epsilon_{\alpha\eta}^{*}\frac{\partial}{\partial X_\eta}\left[\frac{\partial^2}{\partial X_\beta \partial X_\beta}\left(\frac{1}{R}\right)\right]
 +B\epsilon_{\eta\lambda}^{*} \frac{\partial^3}{\partial X_\alpha \partial X_\lambda \partial X_\eta}\left[\frac{\partial^2}{\partial X_\beta \partial X_\beta}\left(\frac{1}{R}\right) \right] \nonumber \\
 & + & C\epsilon_{\eta\lambda}^{*} \frac{\partial^3}{\partial X_\alpha \partial X_\lambda \partial X_\eta} \left[\frac{\partial^2}{\partial  X_\beta \partial X_\beta} R\right] \label{lapUc}
\end{eqnarray}
Here we use the following identities:
\begin{eqnarray}
 &  & \frac{\partial^2}{\partial X_\beta \partial X_\beta}\left(\frac{1}{R}\right)=0,\quad  R \neq 0\nonumber \\
 &  & \frac{\partial^2}{\partial  X_\beta \partial X_\beta} R =\frac{2}{R} \label{lapR}
\end{eqnarray}
Thus we obtain from Eqs. (\ref{lapUc}) and (\ref{lapR}):
\begin{eqnarray}
\frac {\partial^2 u^c_\alpha}{\partial X_\beta \partial X_\beta} & = &
2C\epsilon_{\eta\lambda}^{*}\frac {\partial ^3}{\partial  X_\alpha \partial X_\eta \partial X_\lambda}\left(\frac{1}{R}\right)\label{12}
\end{eqnarray}
and
\begin{eqnarray}
\frac{\partial^2 u^c_\gamma}{\partial X_\alpha \partial X_\gamma} & = &
 \frac{\partial^2}{\partial X_\alpha \partial X_\gamma}\left[A\epsilon_{\gamma\eta}^{*}\frac{\partial}{\partial X_{\eta}}\left(\frac{1}{R}\right)
 +B\epsilon_{\eta\lambda}^{*}\frac{\partial^{3}}{\partial X_{\gamma}\partial X_{\eta}\partial X_{\lambda}}\left(\frac{1}{R}\right)
 +C\epsilon_{\eta\lambda}^{*}\frac{\partial^{3}}{\partial X_{\gamma}\partial X_{\eta}\partial X_{\lambda}}R\right]\nonumber \\
 & = & \frac{\partial}{\partial X_{\alpha}}\left[
 A\epsilon_{\gamma\eta}^{*}\frac{\partial^{2}}{\partial X_{\gamma}\partial X_{\eta}}\left(\frac{1}{R}\right)
 +B\epsilon_{\eta\lambda}^{*}\frac{\partial^{4}}{\partial X_{\gamma}\partial X_{\gamma}\partial X_{\eta}\partial X_{\lambda}}\left(\frac{1}{R}\right)
 +C\epsilon_{\eta\lambda}^{*}\frac{\partial^{4}}{\partial X_{\gamma}\partial X_{\gamma}\partial X_{\eta}\partial X_{\lambda}}R \label{13}
 											\right] \ .\nonumber \\
\end{eqnarray}
Noting that the last two terms involve the Laplacian and we employ Eq. (\ref{lapR}) to obtain:
\begin{eqnarray}
\frac{\partial^2 u^c_\gamma}{\partial X_\alpha \partial X_\gamma} & = &
A\epsilon_{\gamma\eta}^{*} \frac{\partial^3}{\partial X_\alpha \partial X_\gamma \partial X_\eta}\left(\frac{1}{R}\right)
+2C\epsilon_{\eta\lambda}^{*}\frac{\partial^3}{\partial X_\alpha \partial X_\eta \partial X_\lambda} \left(\frac{1}{R}\right)\nonumber \\
 & = & \left(A+2C\right)\epsilon_{\gamma\eta}^{*} \frac{\partial^3}{\partial X_\alpha \partial X_\gamma \partial X_\eta}
 			\left(\frac{1}{R}\right) \label{14}
\end{eqnarray}
Using expressions from Eqs. (\ref{12}) and (\ref{14}) in Eq. (\ref{NL}) we have:
\begin{eqnarray}
 &  & \frac{1}{\left(1-2\nu\right)}\left(A+2C\right)\epsilon_{\eta\gamma}^{*} \frac{\partial^3}{\partial X_\alpha \partial X_\gamma \partial X_\eta} \left(\frac{1}{R}\right)
 +2C\epsilon_{\eta\gamma}^{*} \frac {\partial^3}{\partial X_\alpha \partial X_\gamma \partial X_\eta}\left(\frac{1}{R}\right) =0\ ,\nonumber \\
 &  & \Rightarrow\left[A+2C+2(1-2\nu)C\right]\epsilon_{\eta\gamma}^{*} \frac{\partial^3}{\partial X_\alpha \partial X_\gamma \partial X_\eta}\left(\frac{1}{R}\right) =0\ ,\nonumber \\
 &  & \Rightarrow\left[A+2C+2(1-2\nu)C\right]=0\ ,\nonumber \\
 &  & \Rightarrow C=-\frac{A}{4\left(1-\nu\right)} \ . \label {Ccoef}
\end{eqnarray}
We can thus re-write Eq. (\ref{Uc}) in the following form:
\begin{eqnarray}
u_{\alpha}^{c}=
A\epsilon_{\alpha\eta}^{*}\frac{\partial}{\partial X_\eta}\left(\frac{1}{R}\right)
+B\epsilon_{\eta\lambda}^{*}\frac{\partial^3}{\partial X_\alpha \partial X_\eta \partial X_\lambda}\left(\frac{1}{R}\right)
-\frac{A}{4\left(1-\nu\right)}\epsilon_{\eta\lambda}^{*} \frac{\partial^3}{\partial X_\alpha \partial X_\eta \partial X_\lambda} \left(R\right)\label{16}
\end{eqnarray}
Then, the following identities are used:
\begin{eqnarray}
\frac{\partial}{\partial X_{\eta}}\left(\frac{1}{R}\right) & = & -\frac{X_{\eta}}{R^{3}}\nonumber \\
\frac{\partial^3}{\partial X_\alpha \partial X_\eta \partial X_\lambda}\left(\frac{1}{R}\right) & = &
\frac{-15X_{\alpha}X_{\eta}X_{\lambda}+3r^{2}(X_{\alpha}\delta_{\eta\lambda}
		+X_{\eta}\delta_{\lambda\alpha}+X_{\lambda}\delta_{\eta\alpha})}{R^{7}}\nonumber \\
\frac{\partial^3}{\partial X_\alpha \partial X_\eta \partial X_\lambda}\left(R\right) & = &
 \frac{3X_{\alpha}X_{\eta}X_{\lambda}-R^{2}(X_{\alpha}\delta_{\eta\lambda}
 		+X_{\eta}\delta_{\lambda\alpha}+X_{\lambda}\delta_{\eta\alpha})}{R^{5}}\label{pppdR}
\end{eqnarray}

Using the identities (\ref{pppdR}) in Eq. (\ref{16}), one obtains:
\begin{eqnarray}
u_{\alpha}^{c} & = &
 -A\epsilon_{\alpha\eta}^{*}\frac{X_{\eta}}{R^{3}}
 +B\epsilon_{\eta\lambda}^{*} \frac{-15X_{\alpha}X_{\eta}X_{\lambda}+3r^{2}(X_\alpha \delta_{\eta\lambda} +X_{\eta}\delta_{\lambda\alpha}+X_{\lambda}\delta_{\eta\alpha})}{R^{7}}\nonumber \\
 & - & \frac{A}{4\left(1-\nu\right)}\epsilon_{\eta\lambda}^{*}\frac{3X_{\alpha}X_{\eta}X_{\lambda}-R^{2}(X_{\alpha}\delta_{\eta\lambda}+X_{\eta}\delta_{\lambda\alpha}+X_{\lambda}\delta_{\eta\alpha})}{R^{5}}\nonumber \\
 & = & -A\epsilon_{\alpha\eta}^{*}\frac{X_{\eta}}{R^{3}}
+\left[\frac{3B}{R^{5}}+ \frac{A}{4(1-\nu)R^{3}}\right]\epsilon_{\eta\lambda}^{*}\left(X_\alpha \delta_{\eta\lambda}
+X_{\eta} \delta_{\lambda\alpha}+X_{\lambda} \delta_{\eta\alpha}\right)\nonumber \\
 && \qquad -  \left[\frac{15B}{R^{7}}+\frac{3A}{4(1-\nu)R^{5}}\right]\epsilon_{\eta\lambda}^{*}X_{\alpha}X_{\eta}X_{\lambda} \ .
 \label{18}
\end{eqnarray}
The tensor $\epsilon^*_{\alpha \beta}$ is traceless, therefore these expressions can be somewhat simplified:
\begin{eqnarray}
\epsilon_{\eta\lambda}^{*}\left(X_{\alpha}\delta_{\eta\lambda}
		+X_{\eta}\delta_{\lambda\alpha}+X_{\lambda}\delta_{\eta\alpha}\right)
=2\epsilon_{\alpha\eta}^{*}X_{\eta}
\label{19}
\end{eqnarray}
\begin{eqnarray}
u_{\alpha}^{c} & = &
\left[-\frac{A}{R^{3}}+\frac{6B}{R^{5}}+\frac{2A}{4(1-\nu)R^{3}}\right]\epsilon_{\alpha\eta}^{*}X_{\eta}
-\left[\frac{15B}{R^{7}}+\frac{3A}{4(1-\nu)R^{5}}\right]\epsilon_{\eta\lambda}^{*}X_{\alpha}X_{\eta}X_{\lambda}
\label{20}
\end{eqnarray}
We relate the constrained strain $\epsilon_{\alpha\beta}^{c}$ with the eigenstrain $\epsilon_{\alpha\beta}^{*}$ by the fourth-order
Eshelby tensor $S_{\alpha\beta\gamma\delta}$:
\begin{eqnarray}
\epsilon_{\alpha\beta}^{c}=S_{\alpha\beta\gamma\delta}\epsilon_{\alpha\beta}^{*}
\end{eqnarray}
For a particular case of the spherical inclusion the form of $S_{\alpha\beta\gamma\delta}$ is given by \cite{inclusions}:
\begin{eqnarray}
S_{\alpha\beta\gamma\delta}=\frac{5\nu-1}{15(1-\nu)}\delta_{\alpha\beta}\delta_{\gamma \delta}+\frac{4-5\nu}{15(1-\nu)}(\delta_{\alpha\gamma}\delta_{\beta\delta}+\delta_{\alpha\delta}\delta_{\beta\gamma}) \ .
\label{Stensor}
\end{eqnarray}
For the traceless eigenstrain we have inside the inclusion (and  also for $R=a$ ):\\
 \begin{eqnarray}
u_{\alpha}^{c}=\frac{2(4-5\nu)}{15(1-\nu)}\epsilon_{\alpha\eta}^{*}X_{\eta}.
\end{eqnarray}
So, we can find B in terms of A, demanding that the cubic term in Eq. (\ref{20}) should vanish :
\begin{eqnarray}
 &  & \Rightarrow\frac{15B}{a^7}+\frac{3A}{4(1-\nu)a^5}=0 \ ,\nonumber \\
 &  & \Rightarrow\frac{5B}{a^2}=-\frac{A}{4(1-\nu)} \ ,\nonumber \\
 &  & \Rightarrow B=-\frac{Aa^2}{20(1-\nu)} \ .
\end{eqnarray}
Rewriting $u_{\alpha}^{c}$, one obtains:
\begin{eqnarray}
u_{\alpha}^{c} & = &
\frac{A}{2r^{3}(1-\nu)}\left[(2\nu-1)-\frac{3}{5}\frac{a^{2}}{R^{2}}\right]\epsilon_{\alpha\eta}^{*}X_{\eta}
-\frac{3A}{4r^{5}(1-\nu)}\left[1-\frac{a^{2}}{R^{2}}\right]\epsilon_{\eta\lambda}^{*}X_{\alpha}X_{\eta}X_{\lambda}
\ .\label{17}
\end{eqnarray}
For $R=a$ we have the following matching conditions:
\begin{eqnarray}
\frac{2(4-5\nu)}{15(1-\nu)} & = &
\frac{A}{2a^{3}(1-\nu)} \left[(2 \nu-1)-\frac{3}{5}\frac{a^{2}}{a^{2}}\right] \ ,\nonumber \\
 & = & \frac{A}{2a^{3}(1-\nu)} \left[\frac{2(5\nu-4)}{5}\right]\ ,\nonumber \\
\Rightarrow A & = & -\frac{2}{3}a^{3} \ .
\end{eqnarray}
Using the explicit form for $\epsilon^{*}$ from Eq. (\ref{Eshstrain}), one obtains:
\begin{eqnarray}\label{17}
& & \epsilon^{*}_{\alpha\beta}=(2\lambda_n-\lambda_k)n_{\alpha}n_{\eta}+
(2\lambda_k-\lambda_n)k_{\alpha}k_{\eta}-(\lambda_n+\lambda_k)\delta_{\alpha\eta} \ , \nonumber \\
& & \epsilon^{*}_{\alpha\eta}X_{\eta}=\left((2\lambda_n+\lambda_k)\hat{n}_{\alpha}(\hat{\B n}\cdot \B{X}) +
(\lambda_n+2\lambda_k)\hat{k}_{\alpha}(\hat{\B k}\cdot \B{X})
-(\lambda_n+\lambda_k) X_{\alpha} \right)\ , \nonumber \\
& & X_{\eta}\epsilon^{*}_{\eta\lambda}X_{\lambda}=\left( (2\lambda_n+\lambda_k)(\hat{\B n}\cdot \B{X})^2 + (\lambda_n+2\lambda_k)(\hat{\B k}\cdot \B{X})^2-(\lambda_n+\lambda_k)R^2 \right) \ .
\end{eqnarray}
From this we find the final form of $u_{\alpha}^{c}$ that is used in Eq. (\ref{Uc}):
\begin{eqnarray}
u_{\alpha}^{c} & = &
\frac{1}{3(1-\nu)}\frac{a^{3}}{R^{3}}
\left[(1-2\nu)+\frac{3}{5}\frac{a^{2}}{R^{2}}\right]
\left( (2\lambda_n-\lambda_k) \hat{n}_\alpha( \B {\hat{n}} \cdot \B{X}) + (\lambda_n+2\lambda_k) \hat{k}_{\alpha} (\B{\hat{k}} \cdot \B{X}) - (\lambda_n+\lambda_k)X_{\alpha} \right)\nonumber \\
 & + & \frac{1}{2(1-\nu)}\frac{a^{3}}{R^{5}}
\left[1-\frac{a^{2}}{R^{2}}\right]
\left((2\lambda_n-\lambda_k) ( \B {\hat{n}} \cdot \B{X})^{2} +  (\lambda_n-2\lambda_k)(\B{\hat{k}} \cdot \B{X})^2  - (\lambda_n+\lambda_k)\right) X_\alpha \ .
\end{eqnarray}

\subsection{Cartesian components of Eq. \ref{Uc}:}
In the $XYZ$ orthogonal coordinate system, we choose a unit vector $\mathbf{\hat n}$ which makes an angle $\theta$ with the positive $Z-$ axis and whose projection on the $X-Y$ plane makes an angle $\psi$ with the positive direction of the $X$ axis. The cartesian components of $\mathbf{\hat n}$ are $\left(\sin\theta \cos\psi, \sin\theta \sin\psi, \cos\theta\right)$. We choose another unit vector $\mathbf{\hat k}$, which is orthogonal to $\mathbf{\hat n}$ and lies on the plane containing $\mathbf{\hat n}$ and the $Z$ axis. The components of $\mathbf{\hat k}$ are given by $\left(\cos\theta \cos\psi, \cos\theta \sin\psi, -\sin\theta\right)$. Plugging these into Eq.~\ref{Uc}, we obtain:
\begin{eqnarray}
&& u^c_x = \frac{\epsilon^{*}}{3(1-\nu)}\frac{a^{3}}{R^{3}} \left[(1-2\nu)+\frac{3}{5}\frac{a^{2}}{R^{2}}\right]\bigg[\sin\theta \cos\psi\left(X\sin\theta \cos\psi + Y\sin\theta \sin\psi + Z\cos\theta\right) \nonumber \\
&&  - \cos\theta \cos\psi\left(X\cos\theta \cos\psi + Y\cos\theta \sin\psi - Z\sin\theta\right)\bigg] \nonumber \\
&& + \frac{\epsilon^{*}}{2(1-\nu)}\frac{a^{3}}{R^{5}}
\left[1-\frac{a^{2}}{R^{2}}\right]\bigg[\left(X\sin\theta \cos\psi + Y\sin\theta \sin\psi + Z\cos\theta\right)^2 - \left(X\cos\theta \cos\psi + Y\cos\theta \sin\psi - Z\sin\theta\right)^2\bigg]X \nonumber \\
\end{eqnarray}
\begin{equation}
\boxed{
\begin{aligned}
u^c_x &= \frac{\epsilon^{*}}{3(1-\nu)}\frac{a^{3}}{R^{3}} \left[(1-2\nu)+\frac{3}{5}\frac{a^{2}}{R^{2}}\right]\bigg[ -X\cos^2\psi\cos2\theta - \frac{Y}{2}\sin2\psi\cos2\theta + Z\sin2\theta\cos\psi\bigg] \nonumber \\
+& \frac{\epsilon^{*}}{2(1-\nu)}\frac{a^{3}}{R^{5}}
\left[1-\frac{a^{2}}{R^{2}}\right]\bigg[\left(-X^2\cos^2\psi -Y^2\sin^2\psi + Z^2\right)\cos2\theta - \frac{XY}{2}\sin2\psi\cos2\theta + YZ\sin\psi\sin2\theta + XZ\sin2\theta\cos\psi\bigg]X
\end{aligned}
}
\end{equation}
Similarly,
\begin{eqnarray}
&& u^c_y = \frac{\epsilon^{*}}{3(1-\nu)}\frac{a^{3}}{R^{3}} \left[(1-2\nu)+\frac{3}{5}\frac{a^{2}}{R^{2}}\right]\bigg[\sin\theta \sin\psi\left(X\sin\theta \cos\psi + Y\sin\theta \sin\psi + Z\cos\theta\right) \nonumber \\
&&  - \cos\theta \sin\psi\left(X\cos\theta \cos\psi + Y\cos\theta \sin\psi - Z\sin\theta\right)\bigg] \nonumber \\
&& + \frac{\epsilon^{*}}{2(1-\nu)}\frac{a^{3}}{R^{5}}
\left[1-\frac{a^{2}}{R^{2}}\right]\bigg[\left(X\sin\theta \cos\psi + Y\sin\theta \sin\psi + Z\cos\theta\right)^2 - \left(X\cos\theta \cos\psi + Y\cos\theta \sin\psi - Z\sin\theta\right)^2\bigg]Y \nonumber \\
\end{eqnarray}
\begin{equation}
\boxed{
\begin{aligned}
u^c_y &= \frac{\epsilon^{*}}{3(1-\nu)}\frac{a^{3}}{R^{3}} \left[(1-2\nu)+\frac{3}{5}\frac{a^{2}}{R^{2}}\right]\bigg[ -\frac{X}{2}\sin2\psi\cos2\theta - Y\sin^2\psi\cos2\theta + Z\sin2\theta\sin\psi\bigg] \nonumber \\
+& \frac{\epsilon^{*}}{2(1-\nu)}\frac{a^{3}}{R^{5}}
\left[1-\frac{a^{2}}{R^{2}}\right]\bigg[\left(-X^2\cos^2\psi -Y^2\sin^2\psi + Z^2\right)\cos2\theta - \frac{XY}{2}\sin2\psi\cos2\theta + YZ\sin\psi\sin2\theta + XZ\sin2\theta\cos\psi\bigg]Y
\end{aligned}
}
\end{equation}
and
\begin{eqnarray}
&& u^c_z = \frac{\epsilon^{*}}{3(1-\nu)}\frac{a^{3}}{R^{3}} \left[(1-2\nu)+\frac{3}{5}\frac{a^{2}}{R^{2}}\right]\bigg[\cos\theta \left(X\sin\theta \cos\psi + Y\sin\theta \sin\psi + Z\cos\theta\right) \nonumber \\
&&  - \sin\theta \left(X\cos\theta \cos\psi + Y\cos\theta \sin\psi - Z\sin\theta\right)\bigg] \nonumber \\
&& + \frac{\epsilon^{*}}{2(1-\nu)}\frac{a^{3}}{R^{5}}
\left[1-\frac{a^{2}}{R^{2}}\right]\bigg[\left(X\sin\theta \cos\psi + Y\sin\theta \sin\psi + Z\cos\theta\right)^2 - \left(X\cos\theta \cos\psi + Y\cos\theta \sin\psi - Z\sin\theta\right)^2\bigg]Z \nonumber \\
\end{eqnarray}

\begin{eqnarray}
u^c_z & &=\frac{\epsilon^{*}}{3(1-\nu)}\frac{a^{3}}{R^{3}} \left[(1-2\nu)+\frac{3}{5}\frac{a^{2}}{R^{2}}\right]Z
+ \frac{\epsilon^{*}}{2(1-\nu)}\frac{a^{3}}{R^{5}}
\left[1-\frac{a^{2}}{R^{2}}\right]\times\nonumber \\ && \bigg[\left(-X^2\cos^2\psi -Y^2\sin^2\psi + Z^2\right)\cos2\theta - \frac{XY}{2}\sin2\psi\cos2\theta + YZ\sin\psi\sin2\theta + XZ\sin2\theta\cos\psi\bigg]Z
\end{eqnarray}

\section{Analysis of the energy}
\label{energy}
We start with the expression on Eq. (\ref{EofN}):
\begin {equation}
E=\frac{1}{2}\sum_{i=1}^{\C N} \int_{v_0^{(i)}} \sigma^{(i)}_{\alpha \beta} u^{(i)}_{\alpha, \beta} dV
+ \frac{1}{2} \int_{V- \sum_{i=1}^{\C N}v_0^{(i)}}\sigma^{(m)}_{\alpha \beta} u^{(m)}_{\alpha, \beta} dV \ .
\end{equation}
As the stress tensor is independent of $\alpha$, we can use:
\begin{equation}
\tfrac{\partial}{\partial \alpha}(\sigma_{\alpha \beta}u_{\beta})=
\sigma_{\alpha \beta}u_{\alpha, \beta} +\sigma_{\alpha \beta,\alpha}u_{ \beta}=\sigma_{\alpha \beta}u_{\alpha, \beta}
\end{equation}
and obtain:
\begin {equation}
E=\frac{1}{2}\sum_{i=1}^{N} \int_{v_{0}^{(i)}} \tfrac{\partial}{\partial \alpha}(\sigma^{(i)}_{\alpha \beta}u^{(i)}_{\beta}) dV
+\frac{1}{2} \int_{V- \sum_{i=1}^{N}v_{0}^{(i)}}  \tfrac{\partial}{\partial \alpha}(\sigma^{(m)}_{\alpha \beta}u^{(m)}_{\beta}) dV
\end{equation}
Using Gauss theorem, we can replace the volume integrals with surface ones:
\begin {equation}
E=\frac{1}{2}\sum_{i=1}^{N} \int_{S_{0}^{(i)}}\sigma^{(i)}_{\alpha \beta}u^{(i)}_{\beta}\hat{n}^{(i)}_{\alpha} dS
+\frac{1}{2} \int_{S^{\infty}} \sigma^{(m)}_{\alpha \beta}u^{(m)}_{\beta}  \hat{n}^{\infty}_{\alpha} dS
- \frac{1}{2} \int_{S_{0}^{(i)}} \sigma^{(m)}_{\alpha \beta}u^{(m)}_{\beta}  \hat{n}^{(i)}_{\alpha} dS
\end{equation}
where $\hat{n}^{(i)}$ is a unit vector pointing outwards from the i'th inclusion, and  $\hat{n}^{\infty}$ is a unit vector pointing out of the matrix volume . Now we can sum the matrix and inclusion terms together and obtain:
\begin {equation}
E=\frac{1}{2}\sum_{i=1}^{N} \int_{S_{0}^{(i)}} (\sigma^{(i)}_{\alpha \beta}u^{(i)}_{\beta}-  \sigma^{(m)}_{\alpha \beta}u^{(m)}_{\beta} ) \hat{n}^{(i)}_{\alpha} dS
+\frac{1}{2} \int_{S^{\infty}} \sigma^{(m)}_{\alpha \beta}u^{(m)}_{\beta}  \hat{n}^{\infty}_{\alpha} dS
\label{42}
\end{equation}
For each of the terms, we use the following expressions:
\begin{equation}
\begin{aligned}
\epsilon^{(m)}_{\alpha \beta}\left(\B{X} \right)=
\epsilon^{\infty}_{\alpha \beta}\left(\B{X} \right)
+\sum^{N}_{i=1} \epsilon^{(c,i)}_{\alpha \beta}\left(\B{X} \right)
& \qquad
&\epsilon^{(i)}_{\alpha \beta}\left(\B{X} \right)=
\epsilon^{\infty}_{\alpha \beta}\left(\B{X} \right)
+\sum_{j\neq i} \epsilon^{(c,j)}_{\alpha \beta}\left(\B{X} \right)
+\epsilon^{(c,i)}_{\alpha \beta}\left(\B{X} \right)
-\epsilon^{(*,i)}_{\alpha \beta}\left(\B{X} \right)   \\
\sigma^{(m)}_{\alpha \beta}\left(\B{X} \right)=
\sigma^{\infty}_{\alpha \beta}\left(\B{X} \right)+\sum^{N}_{i=1} \sigma^{(c,i)}_{\alpha \beta}\left(\B{X} \right)
& \qquad
&\sigma^{(i)}_{\alpha \beta}\left(\B{X} \right)=
\sigma^{\infty}_{\alpha \beta}\left(\B{X} \right)
+\sum_{j\neq i} \sigma^{(c,j)}_{\alpha \beta}\left(\B{X} \right)
+\sigma^{(c,i)}_{\alpha \beta}\left(\B{X} \right)
-\sigma^{(*,i)}_{\alpha \beta}\left(\B{X} \right)  \\
u^{(m)}_{\alpha}\left(\B{X} \right)=
u^{\infty}_{\alpha}\left(\B{X} \right)
+\sum^{N}_{i=1} u^{(c,i)}_{\alpha }\left(\B{X} \right)
&\qquad
&u^{(i)}_{\alpha }\left(\B{X} \right)=
u^{\infty}_{\alpha }\left(\B{X} \right)
+\sum_{j\neq i} u^{(c,j)}_{\alpha }\left(\B{X} \right)
+u^{(c,i)}_{\alpha \beta}\left(\B{X} \right)
-\epsilon^{(*,i)}_{\alpha \beta}\left(\B{X} \right)  X_{\beta} \\
\end{aligned}
\label{43}
\end{equation}
On the matrix boundary $S^{\infty}$ we have $\epsilon^{(c,i)}=0$ and  $\sigma^{(c,i)}=0$.
Assuming that $\epsilon^{(\infty)}$ and $\sigma^{(\infty)}$  are constant on the  boundary, and using $u^{(\infty)}_{\beta}=\epsilon^{(\infty)}_{\beta \gamma} X_{\gamma}$,
the second integral at (\ref{42}) is reduced to the following form:
\begin {equation}
\frac{1}{2} \int_{S^{\infty}} \sigma^{(m)}_{\alpha \beta}u^{(m)}_{\beta}  \hat{n}^{\infty}_{\alpha}\: dS
=\frac{1}{2} \int_{S^{\infty}} \sigma^{(\infty)}_{\alpha \beta}u^{(\infty)}_{\beta}  \hat{n}^{\infty}_{\alpha}\:dS
=\frac{1}{2} \sigma^{(\infty)}_{\alpha \beta}\epsilon^{(\infty)}_{\beta \gamma}  \int_{S^{\infty}}  X_{\gamma} \hat{n}^{\infty}_{\alpha} \: dS
=\frac{1}{2} \sigma^{(\infty)}_{\alpha \beta}\epsilon^{(\infty)}_{\beta \gamma} V
\end{equation}
Plugging in the expressions from (\ref{43}), we get:
\begin{equation}
E=\frac{1}{2} \sigma^{(\infty)}_{\alpha \beta}\epsilon^{(\infty)}_{\beta \gamma} V
+\frac{1}{2}\sum_{i=1}^{N} \int_{S_{0}^{(i)}} (\sigma^{(i)}_{\alpha \beta}u^{(i)}_{\beta}-  \sigma^{(m)}_{\alpha \beta}u^{(m)}_{\beta} ) \hat{n}^{(i)}_{\alpha} dS
\end{equation}
 To ensure continuity of the force, we demand that on the inclusion boundary surface $\sigma_{\alpha \beta}^{(i)} n^{(i)}=\sigma_{\alpha \beta}^{(m)} n^{(i)}$ . Then, it follows that
\begin{equation}
E=\frac{1}{2} \sigma^{(\infty)}_{\alpha \beta}\epsilon^{(\infty)}_{\beta \gamma} V
+\frac{1}{2}\sum_{i=1}^{N} \int_{S_{0}^{(i)}} \sigma^{(i)}_{\alpha \beta}(u^{(i)}_{\beta} -  u^{(m)}_{\beta} ) \hat{n}^{(i)}_{\alpha} dS
\end{equation}
Summing over $i$ and using the expressions for $u^{(m)}$ and $u^{(i)}$ from (\ref{43})  and  returning to volume integrals, we'll obtain:
\begin{align}
E=& \frac{1}{2} \sigma^{(\infty)}_{\alpha \beta}\epsilon^{(\infty)}_{\beta \gamma} V
+\frac{1}{2}\sum_{i=1}^{N} \int_{S_{0}^{(i)}} \sigma^{(i)}_{\alpha \beta}(-\epsilon^{(*,i)}_{\beta \gamma}X_{\gamma} ) \hat{n}^{(i)}_{\alpha} dS \nonumber \\
=&  \frac{1}{2} \sigma^{(\infty)}_{\alpha \beta}\epsilon^{(\infty)}_{\beta \gamma} V
-\frac{1}{2}\sum_{i=1}^{N} \int_{S_{0}^{(i)}} \sigma^{(i)}_{\alpha \beta}  \hat{n}^{(i)}_{\alpha} \epsilon^{(*,i)}_{\beta \gamma}X_{\gamma}  dS \nonumber \\
\Rightarrow =&  \frac{1}{2} \sigma^{(\infty)}_{\alpha \beta}\epsilon^{(\infty)}_{\beta \gamma} V
-\frac{1}{2}\sum_{i=1}^{N} \epsilon^{(*,i)}_{\beta \gamma} \int_{v_{0}^{(i)}}\tfrac{\partial}{\partial \alpha}( \sigma^{(i)}_{\alpha \beta}  X_{\gamma} ) dv  \\
\Rightarrow =&  \frac{1}{2} \sigma^{(\infty)}_{\alpha \beta}\epsilon^{(\infty)}_{\beta \gamma} V
-\frac{1}{2}\sum_{i=1}^{N} \epsilon^{(*,i)}_{\beta \gamma} \int_{v_{0}^{(i)}} \sigma^{(i)}_{\alpha \beta} \delta_{\alpha \gamma} dv \nonumber \\
\Rightarrow =&  \frac{1}{2} \sigma^{(\infty)}_{\alpha \beta}\epsilon^{(\infty)}_{\beta \gamma} V
-\frac{1}{2}\sum_{i=1}^{N} \epsilon^{(*,i)}_{\beta \alpha} \int_{v_{0}^{(i)}} \sigma^{(i)}_{\alpha \beta} dv \nonumber
\end{align}
where in line 3 we used the fact that $\epsilon^{(*,i)}$ is constant inside the inclusion. \\
If we assume that $R_{ij}\gg a $ (far field approximation), we can estimate the energy as follows:
\begin{align}
E=&  \frac{1}{2} \sigma^{(\infty)}_{\alpha \beta}\epsilon^{(\infty)}_{\beta \gamma} V
-\frac{1}{2}\sum_{i=1}^{N}  v_{0}^{(i)}  \epsilon^{(*,i)}_{\beta \alpha} \sigma^{(i)}_{\alpha \beta}(\B{X})  \nonumber  \\
=&  \frac{1}{2} \sigma^{(\infty)}_{\alpha \beta}\epsilon^{(\infty)}_{\beta \gamma} V
-\frac{1}{2}\sum_{i=1}^{N}  v_{0}^{(i)}  \epsilon^{(*,i)}_{\beta \alpha}
	\left[ \sigma^{\infty}_{\alpha \beta}
	+\sum_{j\neq i} \sigma^{(c,j)}_{\alpha \beta}\left(\B{X}_{ij} \right)
	+\sigma^{(c,i)}_{\alpha \beta}
	-\sigma^{(*,i)}_{\alpha \beta} \right] \nonumber
\end{align}
when only the mixed term depends on the locations of the inclusions. Rearranging the terms, we get our final expression, $E=E_{mat}+E_{\infty}+E_{esh}+E_{int}$:
\begin{equation}
\begin{aligned}E= & \frac{1}{2}\sigma_{\alpha\beta}^{(\infty)}\epsilon_{\alpha\beta}^{(\infty)}V-\frac{1}{2}\sigma_{\alpha\beta}^{(\infty)}\left(\sum_{i=1}^{N}\epsilon_{\alpha\beta}^{(*,i)}v_{0}^{(i)}\right)\\
 & +\frac{1}{2}\sum_{i=1}^{N}(\sigma_{\alpha\beta}^{(*,i)}-\sigma_{\alpha\beta}^{(c,i)})\epsilon_{\alpha\beta}^{(*,i)}v_{0}^{(i)}-\frac{1}{2}\sum_{i=1}^{N}\epsilon_{\alpha\beta}^{(*,i)}v_{0}^{(i)}\left(\sum_{j\neq i}\sigma_{\alpha\beta}^{(c,j)}(\B{X}_{ij})\right)
\end{aligned}
\label{4E}
\end{equation}

\section{Analysis of the $E_{\rm int}$ term}
\label{Eint}

\begin{eqnarray}
E_{int}= &\displaystyle
-\frac{1}{2}\sum_{i=1}^{N}\epsilon_{\alpha\beta}^{(*,i)}v_{0}^{(i)}
\left(\sum_{j\neq i} \sigma_{\alpha\beta}^{(c,j)}( \B{X}_{ij}) \right)
 =-\frac{1}{2} v_{0} \sum_{<i,j>}
 \left(\epsilon_{\alpha\beta}^{(*,i)}\sigma_{\alpha\beta}^{(c,j)}( \B{X}_{ij}) +\epsilon_{\alpha\beta}^{(*,j)}\sigma_{\alpha\beta}^{(c,i)}( \B{X}_{ij})\right)
\end{eqnarray}
Assuming that all inclusions are aligned according to the $E_{\infty}$ constrain, and looking at each pairwise interaction term $ E_{int}^{ij}$ separately, we obtain:
\begin{eqnarray}
 & E_{int}^{ij}= &
-\frac{1}{2} v_{0} \epsilon^{*}(n_{\alpha}^{(i)}n^{(i)}-k_{\alpha}^{(i)}k_{\beta}^{(i)}) \sigma_{\alpha\beta}^{(c,j)}( \B{X}_{ij}) +\epsilon^{*}(n_{\alpha}^{(j)}n_{\eta}^{(j)}- k_{\alpha}^{(j)}k_{\beta}^{(j)}) \sigma_{\alpha\beta}^{(c,i)}( \B{X}_{ij}) \nonumber \\
\end{eqnarray}
Using the expression for $\sigma_{\alpha\eta}^{c}$, we then have:
\begin{align}
E_{int}^{ij} & = &
-\frac{1}{2} v_{0} \epsilon^{*}( & n_{\alpha}^{(i)}n_{\beta}^{(i)}-k_{\alpha}^{(i)}k_{\beta}^{(i)})
(2\mu\epsilon_{\alpha\beta}^{(c,j)}+\lambda\epsilon_{\eta\eta}^{(c,j)}\delta_{\alpha\beta})\nonumber \\
 & = & -\frac{1}{2} v_{0} \epsilon^{*}( & n_{\alpha}^{(i)}n_{\beta}^{(i)}-k_{\alpha}^{(i)}k_{\beta}^{(i)}) \times
 \Biggl[2\mu\frac{a^{3}}{(R_{ij})^{3}}\frac{\epsilon^{*}}{2(1-\nu)}\Biggl( \nonumber \\
 &  &  & \left[-(1-2\nu)-\frac{a^{2}}{(R_{ij})^{2}}\right]
\Biggl\{\frac{(\hat{\B n}^{(j)}\cdot\B{X}^{(ji)})}{(R_{ij})} \left(\frac{\hat{n}_{\alpha}^{(j)}X_{\beta}^{(ji)}}{(R_{ij})} +\frac{\hat{n}_\beta^{(j)} X_{\alpha}^{(ji)}}{(R_{ij})}\right)
 -\frac{(\hat{\B k}^{(j)} \cdot \B {X}^{(ji)})}{(R_{ij})} \left(\frac{\hat{k}_{\alpha}^{(j)}X_{\beta}^{(ji)}}{(R_{ij})} +\frac{\hat{k}_{\beta}^{(j)}X_{\alpha}^{(ji)}}{(R_{ij})} \right)\Biggr\}\nonumber \\
 &  &  & +\biggl[\frac{2(1-2\nu)}{3}-\frac{2}{5}\frac{a^{2}}{(R_{ij})^{2}}\biggr]
(n_{\alpha}n_{\beta}-k_{\alpha}k_{\beta})\nonumber \\
 &  &  & +\left[-5+7\frac{a^{2}}{(R_{ij})^{2}}\right]
\left(\frac{(\hat{\B n}^{(j)} \cdot \B{X}^{(ji)} )^{2}}{(R_{ij})^{2}} -\frac{(\hat{\B k}^{(j)} \cdot\B{X}^{(ji)} )^{2}}{(R_{ij})^{2}} \right) \frac{X_{\alpha}^{(ji)}X_{\beta}^{(ji)}}{(R_{ij})^{2}} \nonumber \\
 &  &  & +\left[1-\frac{a^{2}}{(R_{ij})^{2}}\right]
\Biggl( \left[ \frac{(\hat{\B n}^{(j)} \cdot\B{X}^{(ji)})}{(R_{ij})} \left( \frac{\hat{n}_{\beta}^{(j)}X_{\alpha}^{(ji)}}{(R_{ij})} +\frac{\hat{n}_{\alpha}^{(j)}X_{\beta}^{(ji)}}{(R_{ij})} \right) \right]
-\left[ \frac{(\hat{\B k}^{(j)} \cdot\B{X})}{(R_{ij})} \left( \frac{\hat{k}_{\beta}^{(j)} X_{\alpha}^{(ji)}}{(R_{ij})} +\frac{\hat{k}_{\alpha}^{(j)}X_{\beta}^{(ji)}}{(R_{ij})} \right) \right] \Biggr) \nonumber \\
 &  &  & +\left[1-\frac{a^{2}}{(R_{ij})^{2}}\right]
\left( \frac{(\hat{\B n}^{(j)}\cdot\B{X}^{(ji)})^{2}}{(R_{ij})^{2}} -\frac{(\hat{\B k}^{(j)} \cdot\B{X}^{(ji)} )^{2}}{(R_{ij})^{2}} \right) \delta_{\alpha\beta}\Biggr) \nonumber \\
 &  &  & \qquad\qquad\qquad+\lambda\epsilon_{\eta\eta}^{(c,j)}\delta_{\alpha\beta}\Biggr]+\langle i\leftrightarrow j\rangle
\end{align}
Summation over indexes will give:
\begin{align}
 & \bigl( n_{\alpha}^{(i)}n_{\beta}^{(i)}-k_{\alpha}^{(i)}k_{\beta}^{(i)} \bigr) \delta_{\alpha\beta} + \langle i\leftrightarrow j\rangle = 0  \\
 \nonumber \\
 & \bigl (n_{\alpha}^{(i)}n_{\beta}^{(i)} -k_{\alpha}^{(i)}k_{\beta}^{(i)} \bigr) \times \nonumber \\
 &\qquad \Biggl\{ \frac{(\B{\hat{n}}^{(j)}\cdot\B{X}^{(ji)})}{(R_{ij})} \left( \frac{n_{\alpha}^{(j)}X_{\beta}^{(ji)}}{(R_{ij})} +\frac{n_{\beta}^{(j)}X_{\alpha}^{(ji)}}{(R_{ij})} \right) -\frac{(\B{\hat{k}}^{(j)}\cdot\B{X}^{(ji)})}{(R_{ij})} \left( \frac{k_{\alpha}^{(j)}X_{\beta}^{(ji)}}{(R_{ij})} +\frac{k_{\beta}^{(j)}X_{\alpha}^{(ji)}}{(R_{ij})} \right) \Biggr\} +\langle i\leftrightarrow j\rangle = \nonumber \\
 & 4(\B{\hat{n}}^{(j)}\cdot\B{\hat{r}}^{(ij)})  (\B{\hat{n}}^{(j)}\cdot\B{\hat{n}}^{(i)}) (\B{\hat{n}}^{(i)}\cdot\B{\hat{r}}^{(ij)})
   -4(\B{\hat{k}}^{(j)}\cdot\B{\hat{r}}^{(ij)}) (\B{\hat{k}}^{(j)}\cdot\B{\hat{n}}^{(i)}) (\B{\hat{n}}^{(i)}\cdot\B{\hat{r}}^{(ij)})  \\
  &\qquad -4(\B{\hat{n}}^{(j)}\cdot\B{\hat{r}}^{(ij)}) (\B{\hat{n}}^{(j)}\cdot\B{\hat{k}}^{(i)}) (\B{\hat{k}}^{(i)}\cdot\B{\hat{r}}^{(ij)})
  + 4(\B{\hat{k}}^{(j)}\cdot\B{\hat{r}}^{(ij)}) (\B{\hat{k}}^{(j)}\cdot\B{\hat{k}}^{(i)}) (\B{\hat{k}}^{(i)}\cdot\B{\hat{r}}^{(ij)})\nonumber \\
  \nonumber \\
 & \bigl( n_{\alpha}^{(i)}n_{\beta}^{(i)}-k_{\alpha}^{(i)}k_{\beta}^{(i)}\bigr) \times \bigl(n_{\alpha}^{(j)}n_{\beta}^{(j)} -k_{\alpha}^{(j)}k_{\beta}^{(j)} \bigr) +\langle i\leftrightarrow j\rangle = \nonumber \\
 & \qquad 2(\B{\hat{n}}^{(j)}\cdot\B{\hat{n}}^{(i)})^{2} +2(\B{\hat{k}}^{(j)}\cdot\B{\hat{k}}^{(i)})^{2} -2(\B{\hat{k}}^{(j)}\cdot\B{\hat{n}}^{(i)})^{2} -2(\B{\hat{k}}^{(i)}\cdot\B{\hat{n}}^{(j)})^{2}  \\
 \nonumber \\
 & \bigl( n_{\alpha}^{(i)}n_{\beta}^{(i)}-k_{\alpha}^{(i)}k_{\beta}^{(i)} \bigr) \times \left( \frac{(\B{\hat{n}}^{(j)}\cdot\B{X}^{(ji)})^{2}}{(R_{ij})^{2}} -\frac{(\B{\hat{k}}^{(j)}\cdot\B{X}^{(ji)})^{2}}{(R_{ij})^{2}} \right) \frac{X_{\alpha}^{(ji)}X_{\beta}^{(ji)}}{(R_{ij})^{2}} +\langle i\leftrightarrow j\rangle = \nonumber \\
 &\qquad \qquad  2\left( \frac{(\B{\hat{n}}^{(j)}\cdot\B{X}^{(ji)})^{2}}{(R_{ij})^{2}} -\frac{(\B{\hat{k}}^{(j)}\cdot\B{X}^{(ji)})^{2}}{(R_{ij})^{2}} \right) \times \left(\frac{(\B{\hat{n}}^{(i)}\cdot\B{X}^{(ji)})^{2}}{(R_{ij})^{2}} -\frac{(\B{\hat{k}}^{(i)}\cdot\B{X}^{(ji)})^{2}}{(R_{ij})^{2}} \right) =\nonumber \\
 &  \qquad\qquad2(\B{\hat{n}}^{(j)}\cdot\B{\hat{r}}^{(ij)})^{2} (\B{\hat{n}}^{(i)}\cdot\B{\hat{r}}^{(ij)})^{2}
  +2(\B{\hat{k}}^{(j)}\cdot\B{\hat{r}}^{(ij)})^{2} (\B{\hat{k}}^{(i)}\cdot\B{\hat{r}}^{(ij)})^{2} \\
  & \qquad\qquad-2(\B{\hat{n}}^{(i)}\cdot\B{\hat{r}}^{(ij)})^{2} (\B{\hat{k}}^{(j)}\cdot\B{\hat{r}}^{(ij)})^{2}
  -2(\B{\hat{k}}^{(i)}\cdot\B{\hat{r}}^{(ij)})^{2} (\B{\hat{n}}^{(j)}\cdot\B{\hat{r}}^{(ij)})^{2}\nonumber
\end{align}
Using the relation between Lam\'e parameter (shear modulus) and Young modulus $\mu=\tfrac{\mathcal{E}}{2(1+\nu)}$
 we get the expression for $E_{int}^{ij}$ :
\begin{align}
E_{int}^{ij} & = &
-v_{0}\frac{\mathcal{E}}{(1+\nu)} \frac{a^{3}}{(R_{ij})^{3}}\frac{(\epsilon^{*})^2}{(1-\nu)}\Biggl(
 8\left[\nu-\frac{a^{2}}{(R_{ij})^{2}}\right]&
 \Bigl \{ (\B{\hat{n}}^{(j)}\cdot\B{\hat{r}}^{(ij)})  (\B{\hat{n}}^{(j)}\cdot\B{\hat{n}}^{(i)}) (\B{\hat{n}}^{(i)}\cdot\B{\hat{r}}^{(ij)})  \nonumber \\
  & & &  -(\B{\hat{k}}^{(j)}\cdot\B{\hat{r}}^{(ij)}) (\B{\hat{k}}^{(j)}\cdot\B{\hat{n}}^{(i)}) (\B{\hat{n}}^{(i)}\cdot\B{\hat{r}}^{(ij)})  \nonumber \\
  & & &\qquad -(\B{\hat{n}}^{(j)}\cdot\B{\hat{r}}^{(ij)}) (\B{\hat{n}}^{(j)}\cdot\B{\hat{k}}^{(i)}) (\B{\hat{k}}^{(i)}\cdot\B{\hat{r}}^{(ij)}) \nonumber \\
 & & & + (\B{\hat{k}}^{(j)}\cdot\B{\hat{r}}^{(ij)}) (\B{\hat{k}}^{(j)}\cdot\B{\hat{k}}^{(i)}) (\B{\hat{k}}^{(i)}\cdot\B{\hat{r}}^{(ij)}) \Bigr \}\nonumber \\
 &  &   + 2\biggl[\frac{2(1-2\nu)}{3}-\frac{2}{5}\frac{a^{2}}{(R_{ij})^{2}}\biggr]&
 \Big \{  (\B{\hat{n}}^{(j)}\cdot\B{\hat{n}}^{(i)})^{2} +(\B{\hat{k}}^{(j)}\cdot\B{\hat{k}}^{(i)})^{2} -(\B{\hat{k}}^{(j)}\cdot\B{\hat{n}}^{(i)})^{2} -(\B{\hat{k}}^{(i)}\cdot\B{\hat{n}}^{(j)})^{2}  \Bigr \}
 \nonumber \\
 &  &   + 2\left[-5+7\frac{a^{2}}{(R_{ij})^{2}}\right]&
\Bigl \{(\B{\hat{n}}^{(j)}\cdot\B{\hat{r}}^{(ij)})^{2} (\B{\hat{n}}^{(i)}\cdot\B{\hat{r}}^{(ij)})^{2}
  +(\B{\hat{k}}^{(j)}\cdot\B{\hat{r}}^{(ij)})^{2} (\B{\hat{k}}^{(i)}\cdot\B{\hat{r}}^{(ij)})^{2} \nonumber \\
& & &   -(\B{\hat{n}}^{(i)}\cdot\B{\hat{r}}^{(ij)})^{2} (\B{\hat{k}}^{(j)}\cdot\B{\hat{r}}^{(ij)})^{2}
  -(\B{\hat{k}}^{(i)}\cdot\B{\hat{r}}^{(ij)})^{2} (\B{\hat{n}}^{(j)}\cdot\B{\hat{r}}^{(ij)})^{2} \Bigr \}
    \Bigr) \nonumber \\
\end{align}
After minimizing the $E_{\infty}$ term, we can assume that all $\B{\hat{n}}^{(i)}$ and $\B{\hat{k}}^{(i)} $ are the same, and obtain:
\begin{align}
E_{int}^{ij} & = &
-v_{0}2\mathcal{E}\frac{a^{3}}{(R_{ij})^{3}}\frac{(\epsilon^{*})^2}{(1-\nu^2)}
\Biggl(
 4\left[\nu-\frac{a^{2}}{(R_{ij})^{2}}\right]&
 \Bigl \{ (\B{\hat{n}}\cdot\B{\hat{r}}^{(ij)})^2
   +(\B{\hat{k}}\cdot\B{\hat{r}}^{(ij)})^2 \Bigr \}\nonumber \\
 &  &   + \biggl[\frac{2(1-2\nu)}{3}-\frac{2}{5}\frac{a^{2}}{(R_{ij})^{2}}\biggr]&
 \Big \{ 2 \Bigr \}
 \nonumber \\
 &  &   + \left[-5+7\frac{a^{2}}{(R_{ij})^{2}}\right]&
\Bigl \{\Bigl( (\B{\hat{n}}\cdot\B{\hat{r}}^{(ij)})^{2}
  -(\B{\hat{k}}\cdot\B{\hat{r}}^{(ij)})^{2} \Bigr)^2
 \Bigr \}  \Biggr)\nonumber \\
\end{align}
\\
Denoting $\tilde{x}\equiv (\B{\hat{n}}\cdot\B{\hat{r}}^{(ij)})^2$ and $\tilde{y} \equiv (\B{\hat{k}}\cdot\B{\hat{r}}^{(ij)})^2$ we get the polynomial expression:
\begin {equation}
E_{int}^{ij} = -(A(R_{ij})(x+y)+B((R_{ij}))(x-y)^2+C(R_{ij}))
\end{equation}
As we assumed  $R>>a$, we obtain the minimal energy configuration along the $x=y$ line, with a slightly lower global minimum at  $x=y=\tfrac{1}{\sqrt{2}}$.\\
To calculate the lattice interaction energy, we  now use polar coordinates where the $\hat{x}$ direction is aligned with the $\hat{n}$ direction and obtain $(\B{\hat{n}}\cdot\B{\hat{r}})=\cos(\phi)\sin(\theta)$ and $(\B{\hat{k}}\cdot\B{\hat{r}})=\sin(\phi)\sin(\theta)$. Assuming that all inclusions are placed on the $\phi=\tfrac{\pi}{4}$ plane, we arrive to the following:
\begin{align}
  E_{int}^{ij}(R_{ij},\theta) & =
-8 v_{0}\mathcal{E} \frac{a^{3}}{(R_{ij})^{3}}\frac{(\epsilon^{*})^2}{(1-\nu^2)}
\Biggl(
 \left[\nu-\frac{a^{2}}{(R_{ij})^{2}}\right]
\sin^2(\theta)
   + \biggl[\frac{(1-2\nu)}{3}-\frac{1}{5}\frac{a^{2}}{(R_{ij})^{2}}\biggr]
   \Biggr)
\end{align}
Taking only the leading order terms, one obtains:
 \begin{align}
  E_{int}^{ij}(R_{ij},\theta) & =
-8 v_{0} \mathcal{E} \frac{a^{3}}{(R_{ij})^{3}}\frac{(\epsilon^{*})^2}{(1-\nu^2)}
\Biggl( \nu \sin^2(\theta) + \biggl[\frac{(1-2\nu)}{3}\biggr]  \Biggr)
\end{align}
 Now it is possible to calculate the energy of a given configuration of the elementary plastic events.
 \end{widetext}

\end{document}